\documentclass[preprint,aps,pre,letterpaper,superscriptaddress,longbibliography]{revtex4-1}
\usepackage{amsmath,amssymb,bm}
\usepackage{mathrsfs}
\usepackage{textcomp}
\usepackage{commath}
\usepackage{mathtools}

\usepackage{hyperref}
\hypersetup{
    colorlinks=true,
    linkcolor=blue,
}
\usepackage{natbib}
\usepackage{placeins}
\usepackage[final]{graphicx} 
\usepackage{rotating}
\usepackage{epstopdf}
\usepackage{float}
\usepackage[lofdepth=1,lotdepth]{subfig}
\usepackage[usenames]{color} 

\newcommand{\Ca}{\mathrm{Ca}}

\graphicspath{{figures/}}  

\makeatletter

\newcommand{\numtoRoman}[1]{\expandafter\@slowromancap\romannumeral #1@}
\makeatother

\begin{document}

\title{Dynamics of deformable straight and curved prolate capsules in simple shear flow}

\author{Xiao Zhang}
\affiliation{
Department of Chemical and Biological Engineering\\
University of Wisconsin-Madison, Madison, WI 53706-1691
}
\author{Wilbur A. Lam}
\affiliation{
Wallace H. Coulter Department of Biomedical Engineering\\
Emory University and Georgia Institute of Technology, Atlanta, GA 30332
}
\affiliation{
Department of Pediatrics, Division of Pediatric Hematology/Oncology, Aflac Cancer and Blood Disorders Center of Children's Healthcare of Atlanta\\
Emory University School of Medicine, Atlanta, GA 30322
}
\affiliation{
Winship Cancer Institute\\
Emory University, Atlanta, GA 30322
}
\affiliation{
Parker H. Petit Institute of Bioengineering and Bioscience\\
Georgia Institute of Technology, Atlanta, GA 30332}
\author{Michael D. Graham}\email{Corresponding author. E-mail: mdgraham@wisc.edu}
\affiliation{
Department of Chemical and Biological Engineering\\
University of Wisconsin-Madison, Madison, WI 53706-1691
}
\date{\today}

\begin{abstract}
This work investigates the motion of neutrally-buoyant, slightly deformable straight and curved prolate capsules in unbounded simple shear flow at zero Reynolds number using direct simulations. The curved capsules serve as a model for the typical crescent-shaped sickle red blood cells in sickle cell disease (SCD). The effects of deformability and curvature on the dynamics are revealed. We show that with low deformability, straight prolate spheroidal capsules exhibit tumbling in the shear plane as their unique asymptotically stable orbit. This result contrasts with that for rigid spheroids, where infinitely many neutrally stable Jeffery orbits exist. The dynamics of curved prolate capsules are more complicated due to a combined effect of deformability and curvature. At short times, depending on the initial orientation, slightly deformable curved prolate capsules exhibit either a Jeffery-like motion such as tumbling or kayaking, or a non-Jeffery-like behavior in which the director (end-to-end vector) of the capsule crosses the shear-gradient plane back and forth. At long times, however, a Jeffery-like quasi-periodic orbit is taken regardless of the initial orientation. We further show that the average of the long-time trajectory can be well approximated using the analytical solution for Jeffery orbits with an effective orbit constant $C_{\textnormal{eff}}$ and aspect ratio $\ell_{\textnormal{eff}}$. These parameters are useful for characterizing the dynamics of curved capsules as a function of given deformability and curvature. As the capsule becomes more deformable or curved, $C_{\textnormal{eff}}$ decreases, indicating a shift of the orbit towards log-rolling motion, while $\ell_{\textnormal{eff}}$ increases weakly as the degree of curvature increases but shows negligible dependency on deformability. As cell deformability, cell shape, and cell-cell interactions are all pathologically altered in blood disorders such as SCD, these results will have clear implications on improving our understanding of the pathophysiology of hematologic disease.

\end{abstract}
\maketitle

\section{INTRODUCTION} \label{sec:introduction}
A healthy human red blood cell (RBC) is a fluid-filled biconcave discoid with an incompressible membrane comprised of a fluid lipid bilayer tethered to an elastic spectrin network \cite{Evans1994}. In the past decades, experimental and computational studies have revealed a rich spectrum of dynamical modes for RBCs in shear flow over a vast domain of parameter space in terms of membrane mechanics of RBC and flow conditions \cite{Goldsmith351,Fischer894,Viallat2007PRL,Secomb2007PRL,Viallat2010PRL,FEDOSOV20102215,Fedosov:2011cf,Yazdani2011PRE,Sinha:2015wt}. In pathological cases where normal blood flow is altered, the dynamics of diseased RBCs can be very different from those of healthy ones because of differences in cell properties. In sickle cell disease (SCD), for example, the intracellular polymerization of sickle hemoglobin (HbS) leads to an abnormality in the shape of individual sickle RBCs, but a large variation in cell morphology has been revealed within the population of sample cells. Byun \emph{et al.} \cite{BYUN20124130} classified the sickle RBCs into echinocytes, discocyte, and the typical crescent-shaped irreversibly sickled cells (ISCs) based on their morphological features. In addition, a general decrease in deformability of sickle RBCs compared to that of normal RBCs has been determined using various techniques for the characterization of cell membrane mechanics;  the extent of membrane stiffening varies greatly depending on the states and conditions of individual sickle RBCs \cite{BYUN20124130,messer1970,Nash73,Evans1443}. The decrease in membrane deformability for non-ISCs (namely, all sickle RBCs other than ISCs \cite{BERTLES884}) is found to be slight under oxygenated conditions \cite{Nash73}, but substantial upon deoxygenation \cite{messer1970}, while ISCs exhibit much lower deformability than non-ISCs in both states \cite{messer1970,Nash73,Evans1443}. All these variations for individual sickle RBCs mentioned above pose challenges to understanding the dynamics of sickle RBCs in flow using computational approaches. In this paper, we perform a systematic investigation of the dynamics of single stiff, fluid-filled curved prolate capsules as a model for the typical crescent-shaped sickle RBCs in the microcirculation, and show the effects of initial orientation, membrane deformability and curvature of the capsules on their dynamical motions.

The model we propose for sickle RBCs in this work derives from spheroidal capsules, the dynamics of which have been of broad interest. The motion of isolated neutrally buoyant rigid spheroidal particles in simple shear flow at zero Reynolds number can be described by trajectories known as Jeffery orbits \cite{Jeffery:1922wb}. During flow the center-of-mass position of the spheroid remains on its initial streamline with no drift in any direction, while its major axis of symmetry traces out a closed periodic orbit depending on its initial orientation. Each orbit corresponds to a specific non-negative orbit constant $C$, so the quantity $C_b = C/(1+C)$ is bounded by zero and one. When $C=C_b = 0$, the particle rolls with its major axis aligned with the vorticity direction, while $C_b = 1 (C=\infty)$ corresponds to a pure tumbling motion in the shear plane. For any intermediate value of $C_b$, the major axis of the particle undergoes a so-called kayaking motion. Bretherton \cite{bretherton_1962} later showed that Jeffery's solution can be applied to describe the dynamics of any axisymmetric rigid body with fore-aft symmetry.

Generalizing Jeffery's work, Hinch and Leal \cite{hinch_leal_1979} derived the equations for the motion of general (non-axisymmetric) ellipsoids in simple shear flow at low Reynolds numbers in terms of Euler angles, and found that the motion of a triaxial ellipsoid with comparable axes follows a doubly periodic pattern. Yarin \emph{et al}. \cite{Yarin:1997cw} numerically determined a chaotic behavior for triaxial ellipsoids with one axis significantly longer than the other two, and proposed an analytical theory to explain the onset of this chaotic motion. Attempts have also been made to examine the behaviors of spheroidal capsules when the assumptions on which Jeffery's theory was based are not strictly satisfied. Dupont \emph{et al.} \cite{Dupont2013} and Cordasco and Bagchi \cite{Cordasco:2013hb} recently explored the effects of membrane deformability on the dynamics of fluid-filled spheroidal capsules. Through a stability investigation, Dupont \emph{et al.} \cite{Dupont2013} showed that regardless of the initial orientation, the dynamics of prolate spheroids in Stokes flow converge towards rolling (with the membrane tank-treading), wobbling, and oscillating-swinging regimes, respectively, as deformability increases. The model for the membrane mechanics in this work describes shear elasticity and area dilatation while neglecting bending elasticity. Cordasco and Bagchi \cite{Cordasco:2013hb} performed simulations over a range of parameters for spheroidal capsules using a membrane model that incorporates shear elasticity, area dilatation and bending rigidity, and observed that unlike rigid ellipsoids in Stokes flow, capsules undergo various motions such as precessing and kayaking depending on the initial shape, deformability, and the ratio of the internal to external fluid viscosity. In addition, Mao and Alexeev \cite{mao_alexeev_2014} examined the effects of fluid inertia and particle inertia on the dynamics of single prolate and oblate spheroids at low and moderate Reynolds number using simulations, and revealed that the motion of a single spheroid is dependent on the particle Reynolds number, aspect ratio, initial orientation and the Stokes number \cite{Lundell2010}. 

Other studies have also focused on slender particles such as curved fibers and revealed more complex behaviors. Wang \emph{et al}. \cite{Wang:2012ji} simulated the motion of isolated, rigid, neutrally-buoyant, non-Brownian, slightly curved fibers in simple shear flow at low Reynolds numbers. They revealed that for some initial orientations the fibers drift in the gradient direction resulting from the combined effects of the so-called ``flipping, scooping, and spinning" motions, and the drift rate strongly depends on the initial orientation, aspect ratio, and curvature of the fibers. Crowdy \cite{Crowdy2016} derived a dynamical system governing the motion of a curved rigid two-dimensional circular-arc fiber in simple shear, and the analytical solutions of this dynamical system display the ``flipping" and ``scooping" motions observed in \cite{Wang:2012ji}. Another example of recent works on objects with complex dynamics is that of Dutta and Graham \cite{dutta2017}. They investigated the dynamics in shear flow of inertialess Miura-patterned foldable sheets containing predefined crease lines, and found a rich spectrum of quasi-periodic or periodic motions depending on the initial configuration of the sheet. 
  
In this paper, we investigate the dynamics of single straight and curved prolate  capsules in unbounded simple shear flow at zero Reynolds number and discuss the effects of initial orientation, membrane deformability, and degree of curvature on their dynamics. The rest of the paper is organized as follows: in Section \ref{sec:methods} we present the model for sickle RBCs and the numerical algorithm employed to obtain the fluid field; in Section \ref{sec:results} we first provide a validation of our numerical simulation, followed by detailed results and discussion on the dynamics of single straight and curved prolate capsules; lastly, concluding remarks are presented in Section \ref{sec:conclusion}.

\section{MODEL FORMULATION} \label{sec:methods}
    \subsection{Model and discretization} \label{sec:sickle_model}
  \begin{figure}[h]
  \centering
  \captionsetup{justification=raggedright}
  \includegraphics[width=0.4\textwidth]{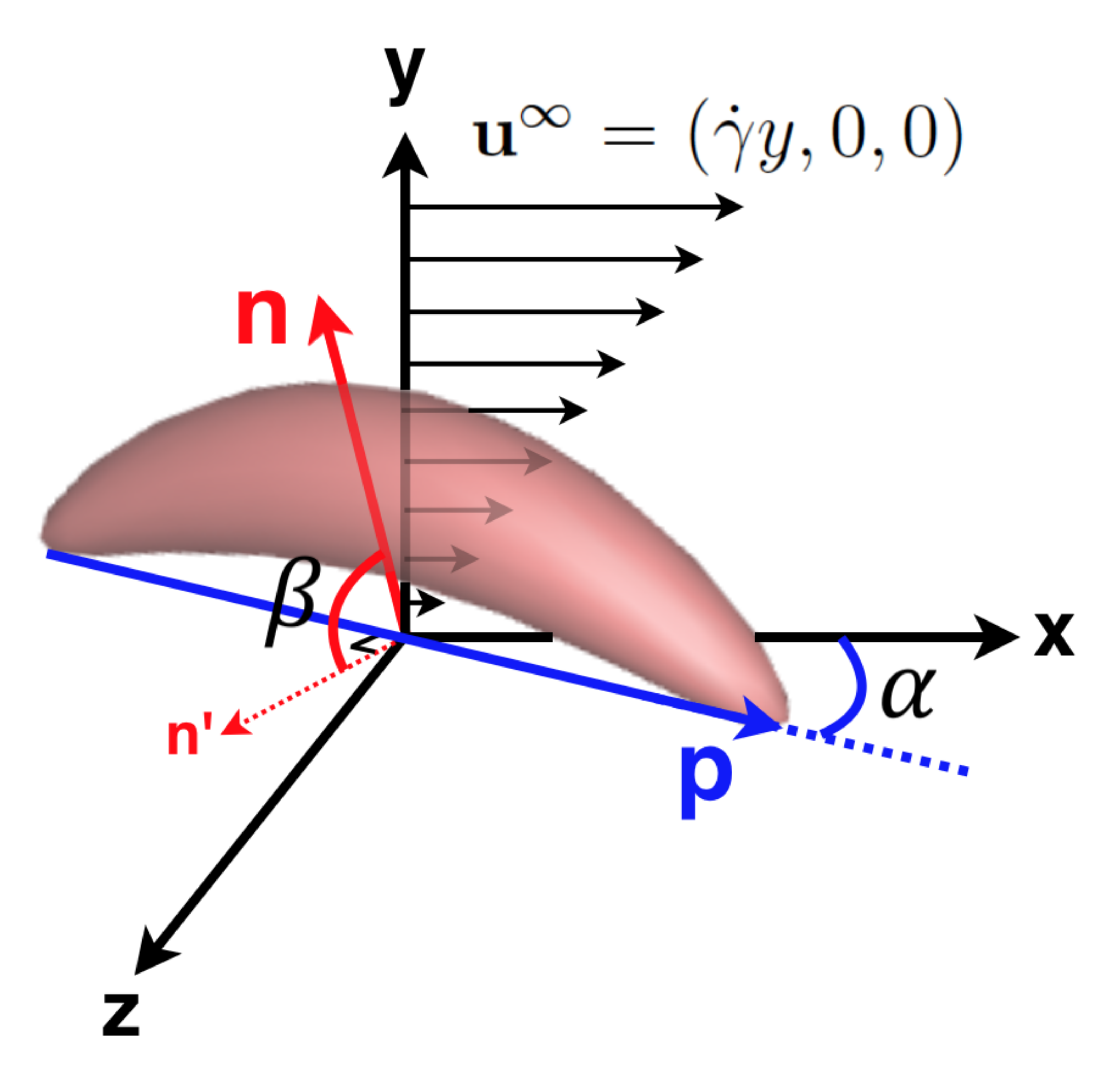}
  \caption[determination of orientation]{Schematic of the 3D orientation of a curved prolate  capsule in unbounded simple shear flow.}
  \label{fig:orientation_determination}
  \end{figure}

We consider an isolated neutrally-buoyant fluid-filled deformable capsule with a straight or curved prolate  rest shape in unbounded simple shear flow (FIG.~\ref{fig:orientation_determination}). The shear rate is $\dot{\gamma}$, and the undisturbed flow velocity is given as $\mathbf{u}^{\infty} = (\dot{\gamma} y, 0, 0)$. Both the suspending fluid and the fluid inside the capsule are assumed to be incompressible and Newtonian with the same viscosity. The prolate spheroidal shape of the capsule derives from polar stretching and equatorial compression of a spherical capsule with radius $s$ centered at the origin. The equation for this transformation is given by
\begin{equation} \label{eq:prolate_equation}
  \mathbf{x}^p = 
\begin{pmatrix} 
  c/s & 0 & 0 \\ 
  0 & a/s & 0 \\
  0 & 0 & a/s 
\end{pmatrix}
  \cdot \mathbf{x}^s,
\end{equation}
where $\mathbf{x}^s$ and $\mathbf{x}^p$ are the positions of a point on the membrane of the original spherical capsule and the mapped point on the membrane of the resulting prolate spheroidal capsule with a polar radius $c$ and an equatorial radius $a$ (aspect ratio $r_p = c/a$). The resulting prolate spheroidal capsule is assumed to align with the $x$ axis. To generate a curved capsule, we then impose a quadratic unidirectional displacement of membrane points perpendicular to the polar axis of the prolate spheroidal capsule (here in the $y$ direction),
\begin{subequations} \label{eq:curvature_equation}
\begin{equation} \label{eq:curvature_equation1}
x^c_1 = x^p_1,
\end{equation}	
\begin{equation} \label{eq:curvature_equation2}
x^c_2 = x^p_2 - k \Big[x^p_1 - \frac{1}{2} (x^{pl}_1 + x^{pr}_1)\Big]^2,
\end{equation}
\begin{equation} \label{eq:curvature_equation3}
x^c_3 = x^p_3, 
\end{equation}
\end{subequations}
where $\mathbf{x}^{pl}$ and $\mathbf{x}^{pr}$ are the positions of the two ends (left- and right-most membrane points) of the prolate spheroid, and $\mathbf{x}^c = (x^c_1,x^c_2,x^c_3)$ is the position of a membrane point of the resulting curved prolate capsule mapped with the membrane point of the prolate spheroid at position $\mathbf{x}^p = (x^p_1,x^p_2,x^p_3)$; $k$ has units of inverse length, which can be nondimensionalized by $c$ to define the degree of curvature $K = k c$.

To describe the behavior of the capsule membrane in response to the in-plane shear elastic force, a membrane model by Skalak \emph{et al.} \cite{SKALAK:1973tp} is utilized, in which the strain energy density $W$ is given by
\begin{equation} \label{eq:Skalak_model}
W_{\mathrm{SK}} = \frac{G}{4}[(I_1^2 + 2 I_1 - 2 I_2) + C_a I_2^2],
\end{equation}	
where $G$ is the in-plane shear modulus of the membrane, and $C_a$ is a material property that characterizes the energy penalty for area change of the membrane; the strain invariants $I_1$ and $I_2$ are functions of the principal stretch ratios $\lambda_1$ and $\lambda_2$, defined as
\begin{equation} \label{eq:strain_invariants}
I_1 = \lambda_1^2 + \lambda_2^2 - 2, \quad I_2 = \lambda_1^2 \lambda_2^2 - 1.
\end{equation} 
Note that this model predicts a strain hardening behavior of the membrane, consistent with the experimentally determined response of an RBC membrane to stretching \cite{Sinha:2015wt,Mills2004}. The deformability of the capsule is characterized by the nondimensional capillary number Ca = $ \eta \dot{\gamma} c/G$, where $c$ again is half-length of the straight or curved prolate capsule. In this work the capsules are taken to be stiff in accordance with the biomechanical properties of the membrane of a sickle RBC, with Ca lower than 0.2 assuming the shear rate $\dot{\gamma}$ of blood flow in the microcirculation $\sim 10 - 10^3$ s$^{-1}$ \cite{Lipowsky2013}, the characteristic length $c$ of a typical sickle RBC $\sim 5 - 6$ $\mu$m \cite{BYUN20124130}, the viscosity of plasma $\eta \sim$ 1.71 mPa s \cite{Laogun1980}, and the in-plane shear modulus of a sickle RBC membrane $G \sim 29.8$ $\mu$N/m \cite{BYUN20124130}. The bending modulus of the capsule $K_B$ is expressed nondimensionally by $\hat{\kappa}_B = K_B/c^2 G$, which is $O(10^{-4} - 10^{-2})$ in the physiological context \cite{Evans2008,Betz15320}; here we set $\hat{\kappa}_B = 0.04$ for all capsules. In this work, the natural (spontaneous) shapes for shear and bending elasticities of the capsule membrane are both chosen to be the same as its rest shape, so that any in-plane or out-of-plane deformation would lead to an increase in the membrane energy and an elastic restoring force. The capsule membrane is discretized into 320 piecewise flat triangular elements, resulting in 162 nodes. We have verified that increasing the number of nodes makes no difference to the cell dynamics. Based on this discretization, the calculation of the membrane force density follows the work of Kumar and Graham \cite{Kumar:2012ev} and Sinha and Graham \cite{Sinha:2015wt} using approaches given by Charrier \emph{et al.} \cite{charrier1989free} for the in-plane shear force density and Meyer \emph{et al.} \cite{Meyer:2002vh} for the out-of-plane bending force density, respectively. 

The orientation of a curved capsule is defined by the normalized end-to-end vector \textbf{p} and the unit normal vector \textbf{n}, as illustrated in FIG. ~\ref{fig:orientation_determination}. In this work, initial conditions are chosen so that \textbf{p} is initially in the $x$-$z$ plane, and the initial angle between \textbf{p} and the $x$ axis is denoted as $\alpha$. The initial angle between \textbf{n} and its projection $\mathbf{n'}$ onto the $x$-$z$ plane is $\beta$ (note that $\mathbf{n'} \perp \mathbf{p}$). Thus the initial orientation of the capsule is given by $\alpha$ and $\beta$. For example, ($\alpha, \beta$) = (0,$\pi/2$) means that the capsule is initially in the $x$-$y$ plane with \textbf{p} aligned with the $x$ axis and \textbf{n} with the $y$ axis. If the capsule is not curved, then $\beta$ is irrelevant to its orientation and dynamics. 
    \subsection{Fluid motion}
As noted in Section~\ref{sec:sickle_model}, the characteristic shear rate $\dot{\gamma}$ of blood flow in the microcirculation is $\sim$ $10 - 10^3$ s$^{-1}$ \cite{Lipowsky2013}, the length scale $c$ of a typical sickle RBC $\sim 5 - 6$ $\mu$m \cite{BYUN20124130}, the viscosity of plasma $\eta$ $\sim$ 1.71 mPa s \cite{Laogun1980}, and its density $\rho$ $\sim$ $10^3$ kg/$\textrm{m}^3$. The magnitude of the particle Reynolds number $\textnormal{Re}_p = \rho \dot{\gamma} c^2/\eta$ obtained using these parameters is $\sim O(10^{-4} - 10^{-2})$, which is assumed to be sufficiently small so that the fluid motion is governed by the Stokes equation. To determine the velocity field at each time instant, we employ a boundary integral method \cite{pozrikidis1992boundary,Kumar:2012ev} for simulation in unbounded simple shear flow. Once the flow field is obtained, the positions of the element nodes on the discretized capsule membrane are advanced using a second-order Adams-Bashforth method with time step $\Delta t = 0.02$Ca$l$, where $l$ is the minimum node-to-node distance. In this work, the midpoint of the \textbf{p} vector of the capsule is initially placed at the origin of the unbounded domain. 
\section{RESULTS AND DISCUSSION} \label{sec:results}
   \subsection{Model validation} \label{sec:model_validation}
\begin{figure}[h]
\centering
\captionsetup{justification=raggedright}
 \subfloat[]
{
    \includegraphics[width=0.42\textwidth]{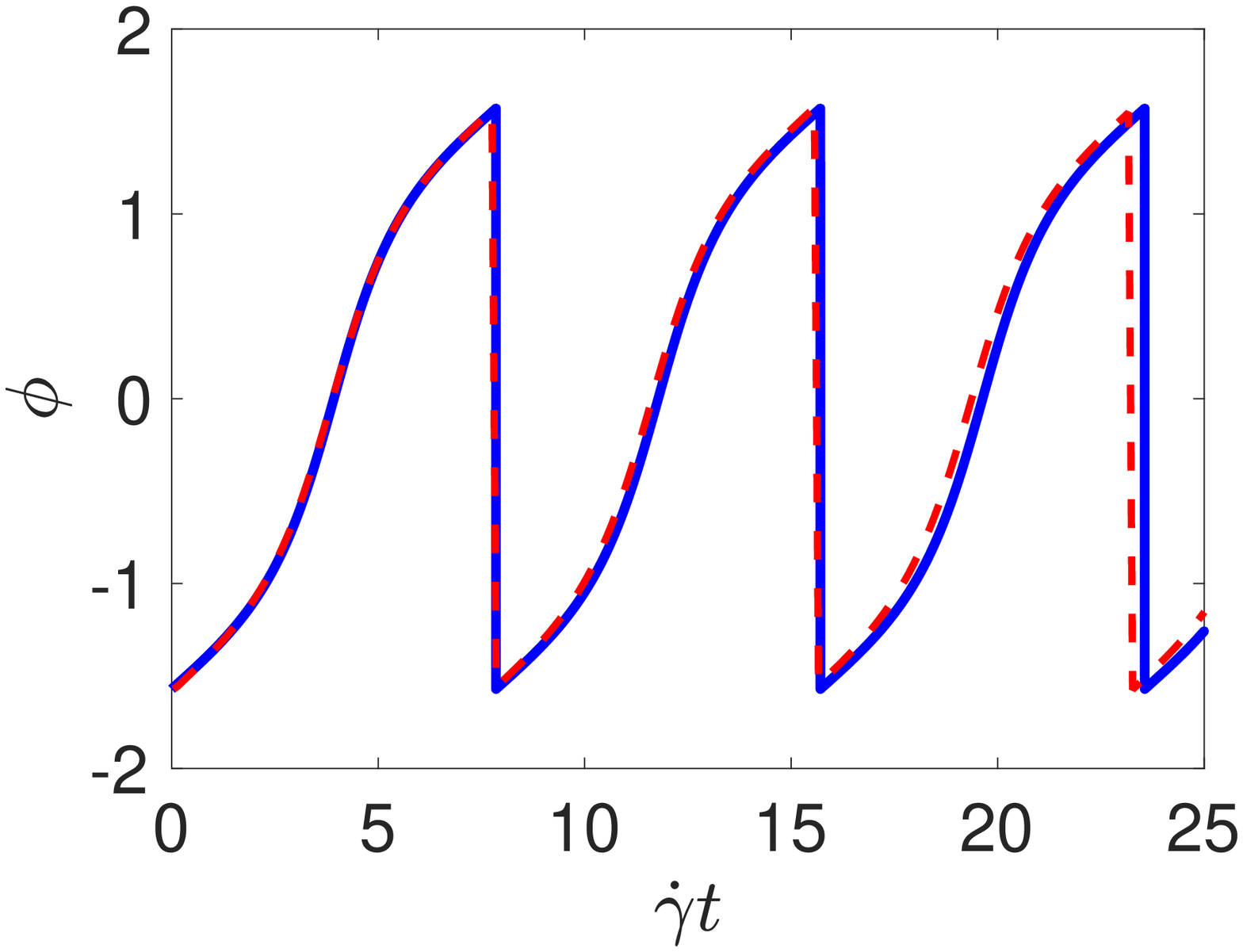}
    \label{fig:model_validation_AR_small}
}
\subfloat[] 
{
    \includegraphics[width=0.42\textwidth]{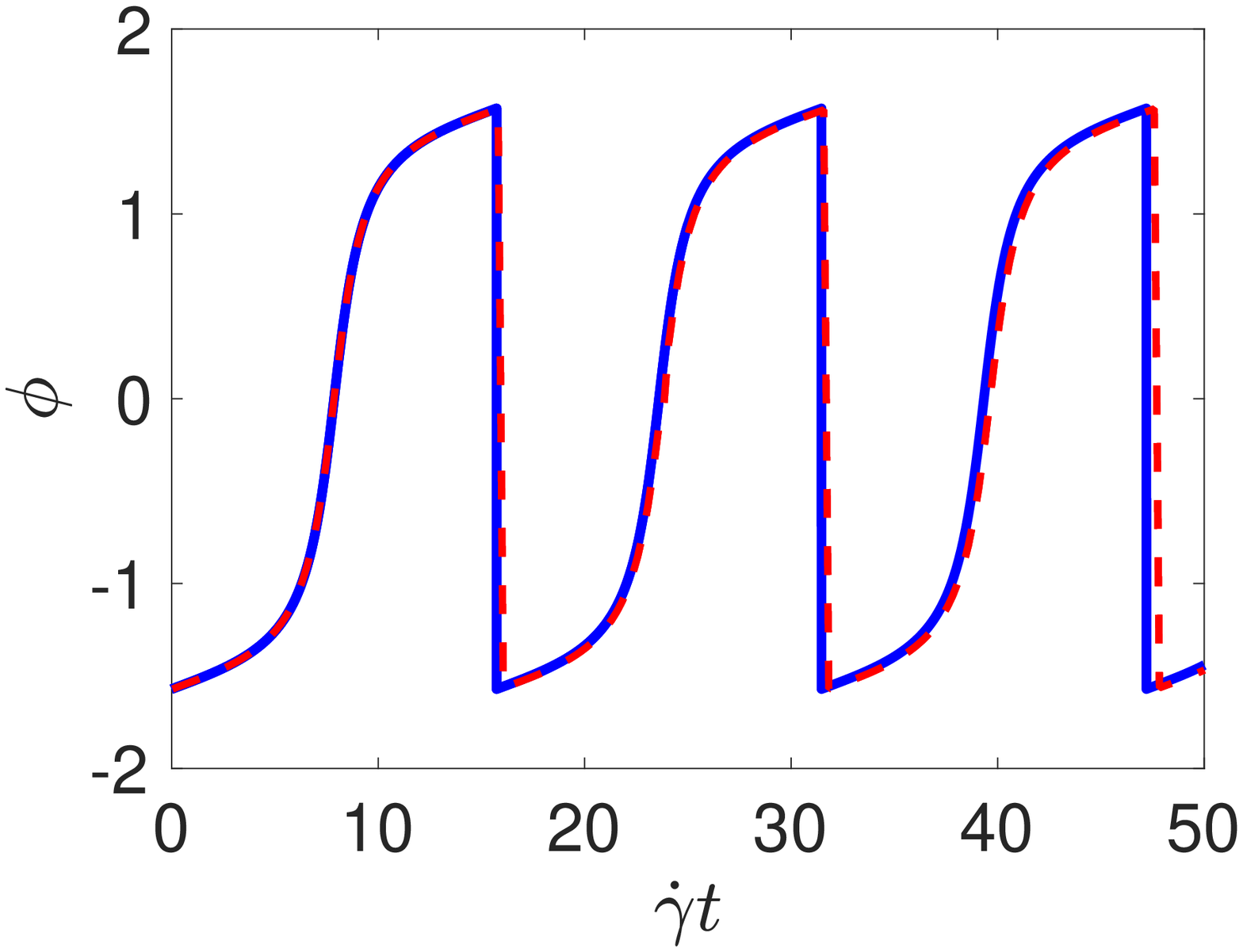}
    \label{fig:model_validation_AR_large}
}
\caption[model_validation]{Evolution of inclination angle $\phi$ for a stiff prolate spheroid (Ca = 0.06) with (a) $r_p$ = 2.0 and (b) $r_p$ = 4.8, respectively. Solid blue lines are predictions by Jeffery's theory and red dashed lines are simulation results.}
\label{fig:model_validation}
\end{figure}

To validate our model, we compare the numerically determined inclination profile for a stiff prolate spheroid (Ca = 0.06) with the prediction by Jeffery's theory \cite{Jeffery:1922wb}, 
\begin{equation} \label{eq:Jeffery_orbit_1}
  \mathrm{tan} \; \phi = r_p \mathrm{tan}\bigg(\frac{\dot{\gamma}t}{r_p+{r_p}^{-1}}+\phi_0\bigg),
\end{equation}
where $\phi$ is the inclination angle between the major axis of the prolate spheroid and the $y$ direction, and $\phi_0$ its initial value; here $\phi_0 = \pi/2$ since the prolate spheroid is initially aligned with the $x$ axis. FIG.~\ref{fig:model_validation} shows the evolution of $\phi$ with nondimensionalized time $\dot{\gamma}t$. A satisfactory agreement between numerical simulation and Jeffery's theory (Eq.~\ref{eq:Jeffery_orbit_1}) is observed for the prolate spheroid with either a low ($r_p$ = 2.0) or a high ($r_p$ = 4.8) aspect ratio. The very small discrepancy in each case derives from the non-zero deformability of the prolate spheroid, and vanishes as Ca approaches zero.   

   \subsection{Dynamics of deformable prolate spheroids} 
      \label{sec:unbounded_non_curved}
\begin{figure}[h]
\centering
\captionsetup{justification=raggedright}
 \subfloat[]
{
    \includegraphics[width=0.40\textwidth]{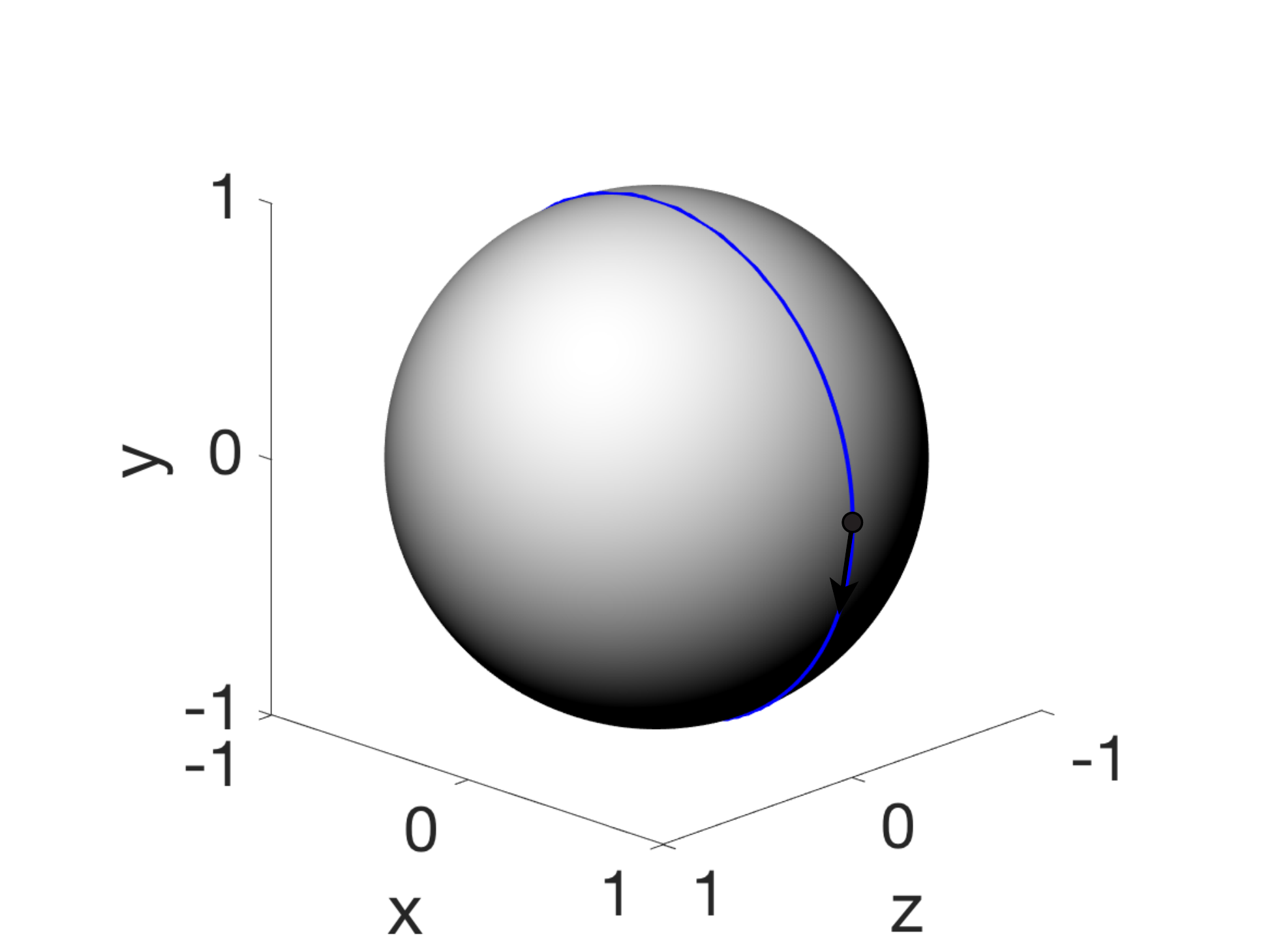}
    \label{fig:unbounded_0_control_Ca_0.1_p}
}
\subfloat[] 
{
    \includegraphics[width=0.40\textwidth]{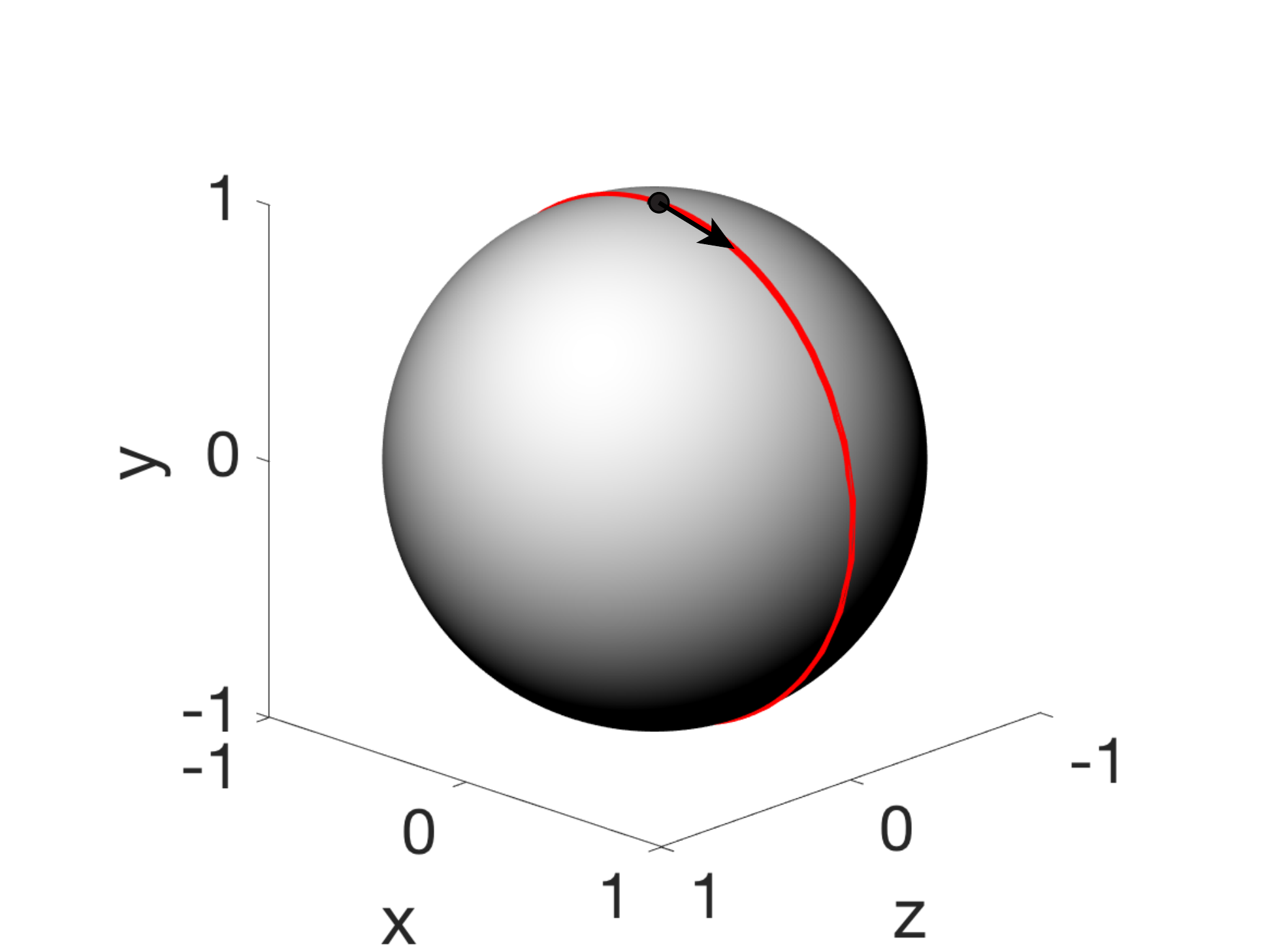}
    \label{fig:unbounded_0_control_Ca_0.1_n}
}
\caption[unbounded_control_0_trajectories]{Trajectories on the unit sphere of (a) \textbf{p} and (b) \textbf{n} for a prolate spheroid ($r_p$ = 4.8, Ca = 0.12) with initial orientation $\alpha = 0$ (up to $\dot{\gamma}t = 2000$).}
\label{unbounded_0_control_Ca_p1_trajectories}
\end{figure}

We first consider a stiff prolate spheroidal capsule ($r_p$ = 4.8, Ca = 0.12) initially aligned with the $x$ axis, $i.e.$, $\alpha = 0$. The long-time trajectories (up to $\dot{\gamma}t$ = 2000) of the end-to-end vector \textbf{p} and the normal vector \textbf{n} on the unit sphere are plotted in FIG.~\ref{unbounded_0_control_Ca_p1_trajectories}. In this case, both \textbf{p} and \textbf{n} appear to trace out a closed circle, indicating that the capsule is taking a steady symmetric tumbling motion in the $x$-$y$ plane. 

\begin{figure}[h]
\centering
\captionsetup{justification=raggedright}
\subfloat[] 
{
    \includegraphics[width=0.40\textwidth]{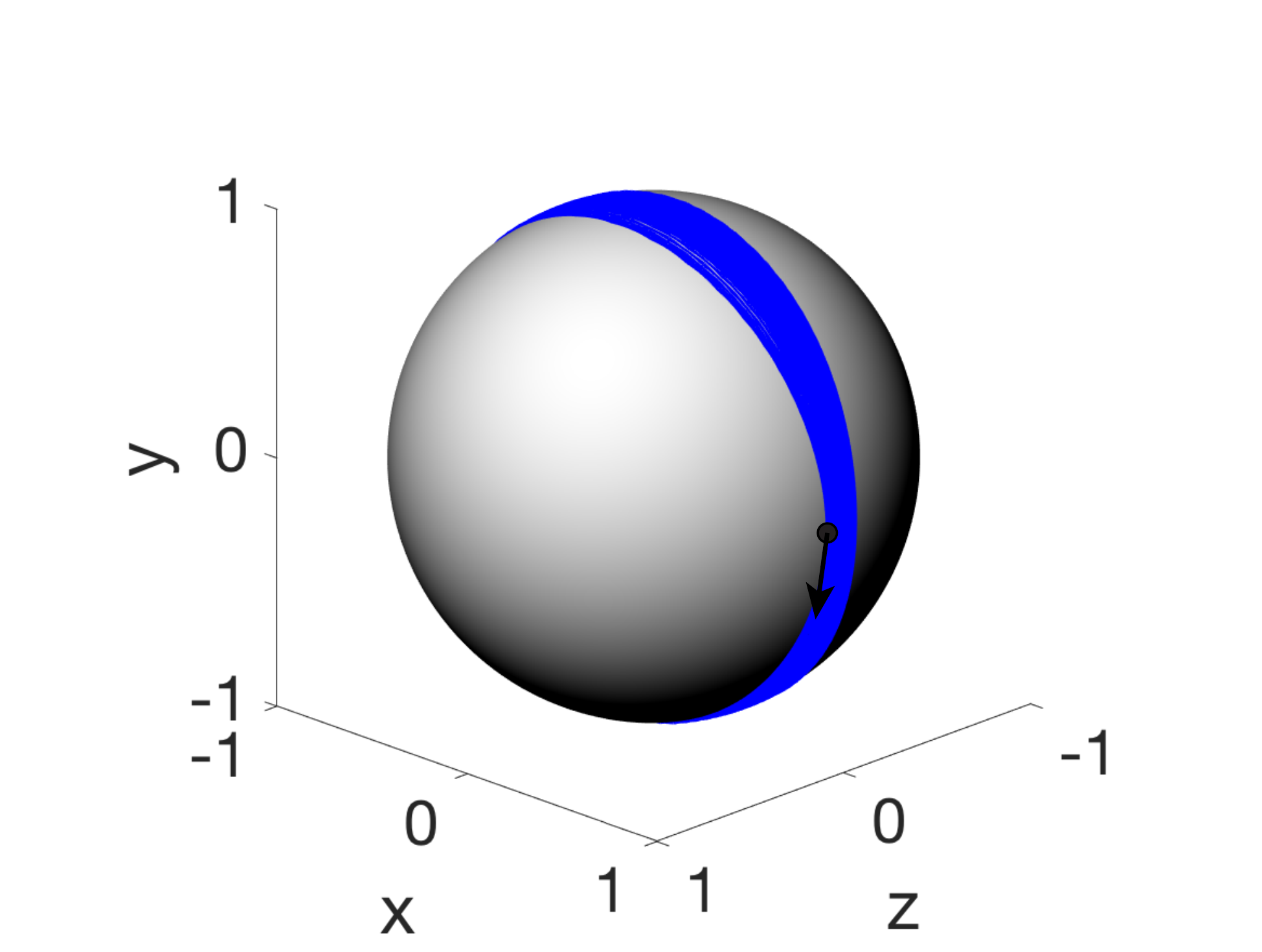}
    \label{fig:unbounded_5_control_p}
}
\subfloat[] 
{
    \includegraphics[width=0.34\textwidth]{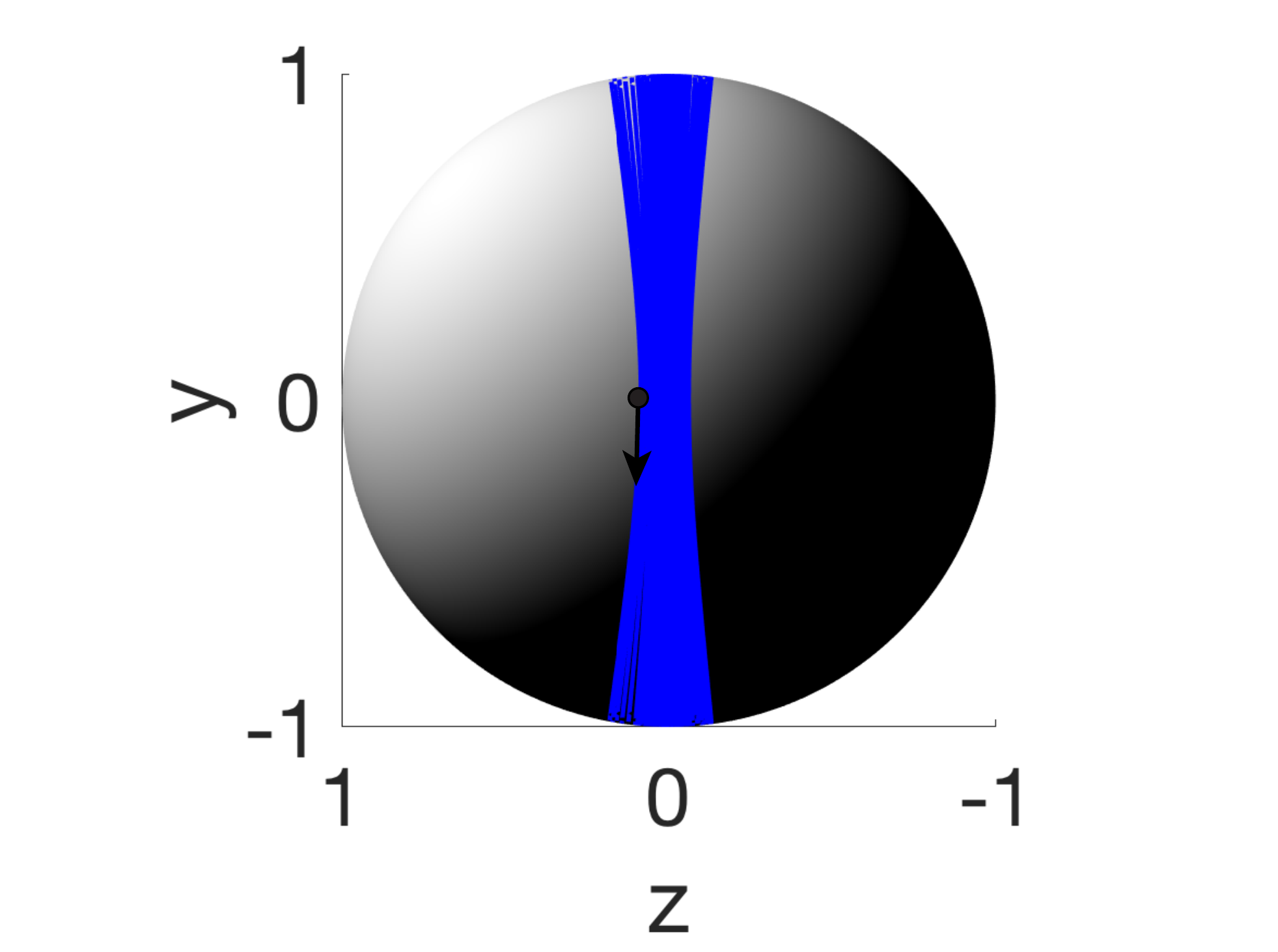}
    \label{fig:unbounded_5_control_p_front}
}
\\
\subfloat[] 
{
    \includegraphics[width=0.40\textwidth]{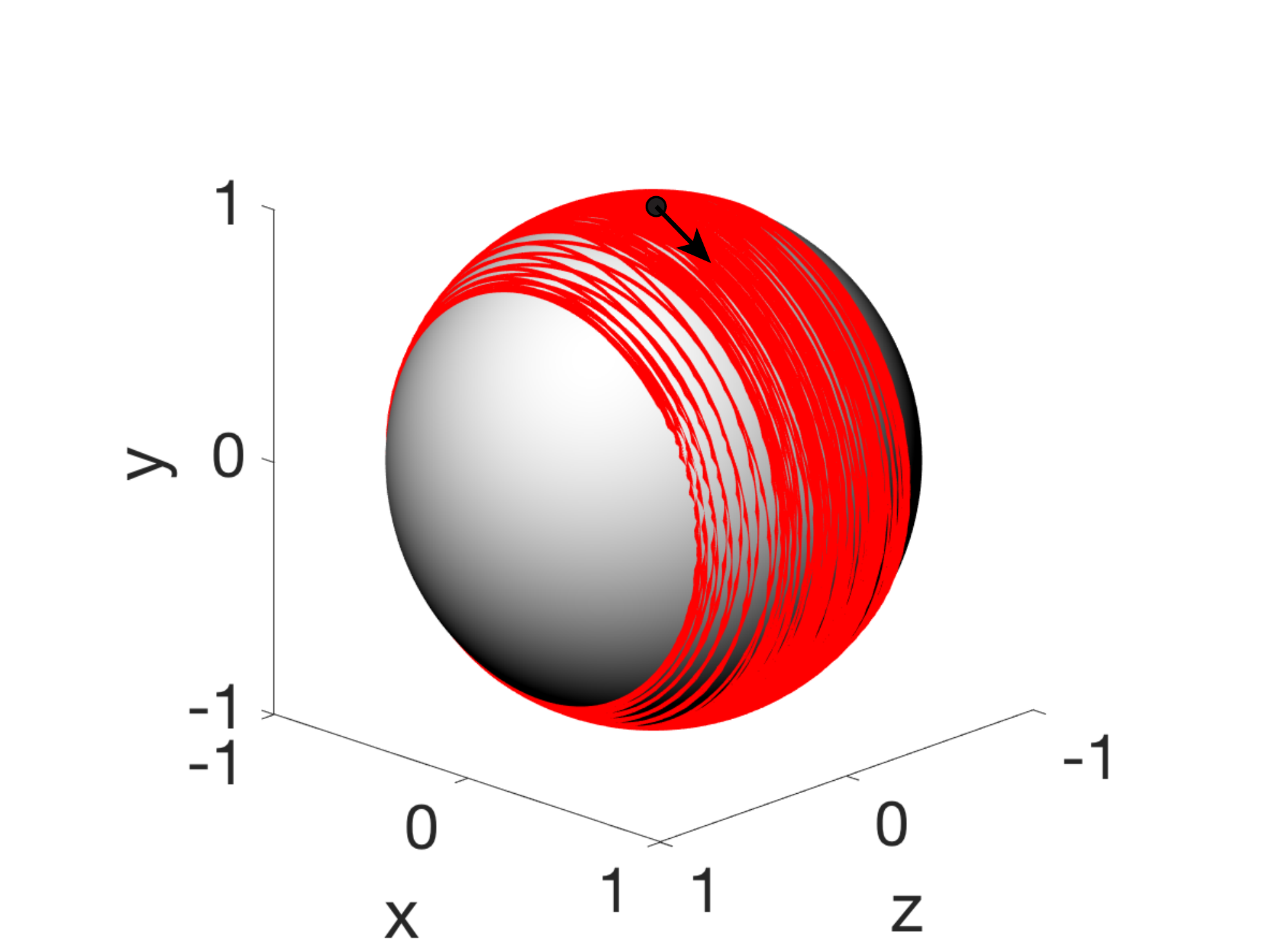}
    \label{fig:unbounded_5_control_n}
}
\subfloat[] 
{
    \includegraphics[width=0.34\textwidth]{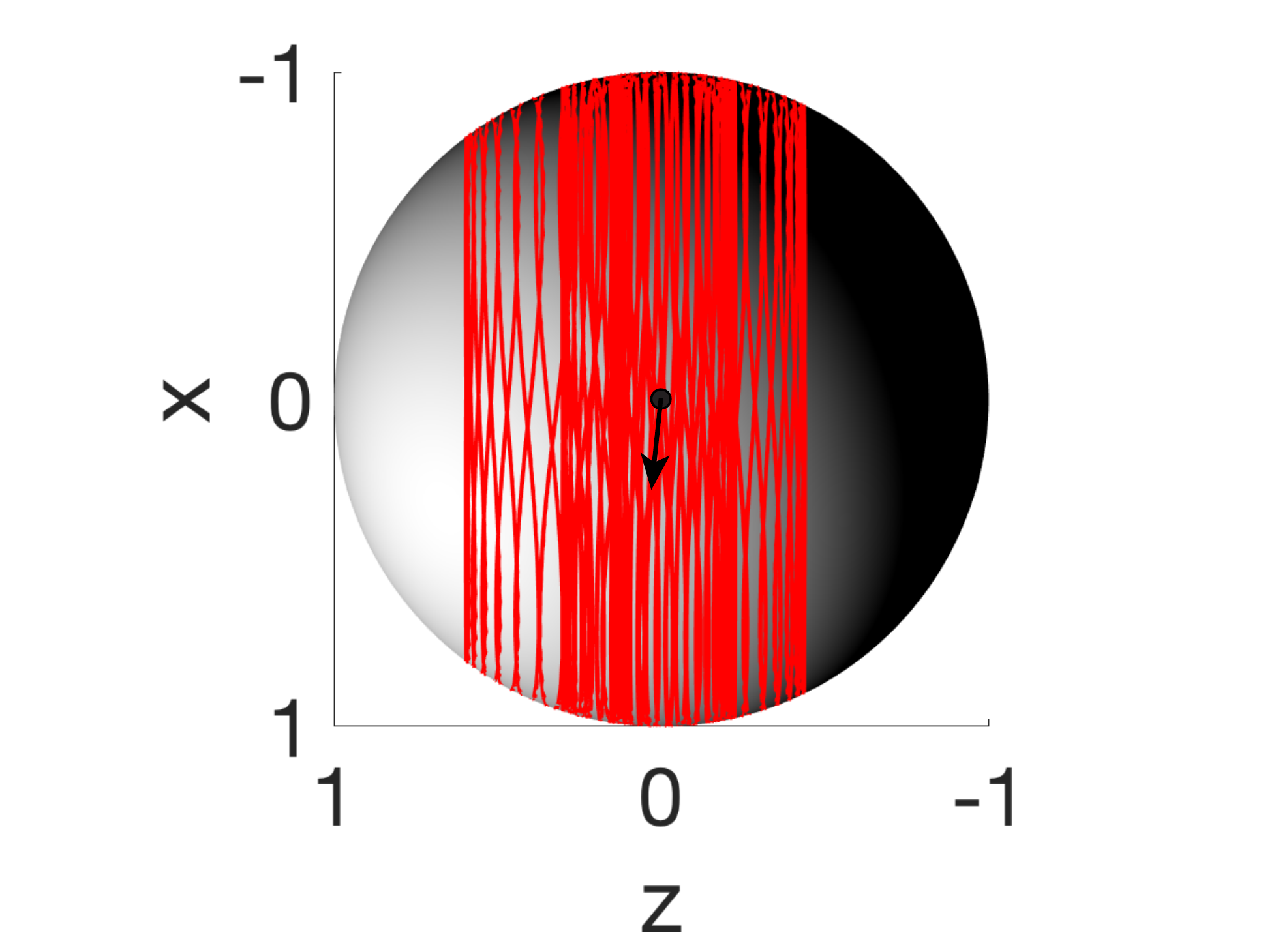}
    \label{fig:unbounded_5_control_n_top}
}
\caption[unbounded_5_control_p_n]{Trajectories on the unit sphere of \textbf{p} ((a) and (b)) and \textbf{n} ((c) and (d)) for a prolate spheroid ($r_p$ = 4.8, Ca = 0.12) with initial orientation $\alpha = \pi/36$.}
\label{unbounded_5_control}
\end{figure}

\begin{figure}[h]
\centering
\captionsetup{justification=raggedright}
{
    \includegraphics[width=0.65\textwidth]{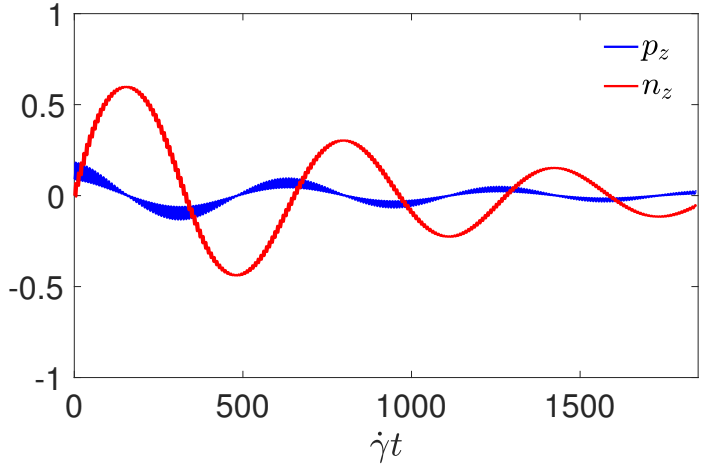}
}
\caption[]{The evolution of the $p_z$ and $n_z$ for a prolate spheroid ($r_p$ = 4.8, Ca = 0.12) with initial orientation $\alpha = \pi/36$.}
\label{fig:unbounded_5_control_pz_nz_t}
\end{figure}

We then consider a perturbation of the initial orientation by a small angle $\alpha = \pi/36$. The trajectories of \textbf{p} and \textbf{n} vectors on the unit sphere are shown in FIG.~\ref{unbounded_5_control}. It can be observed that both $p_z$ and $n_z$ show an oscillation around zero (FIG.~\ref{fig:unbounded_5_control_pz_nz_t}) whose amplitude decays to zero as time increases. The director \textbf{p} crosses the $x$-$y$ plane back-and-forth (as indicated by the sign change of $p_z$ in FIG.~\ref{fig:unbounded_5_control_pz_nz_t}), exhibiting non-Jeffery-like dynamics (note that in Jeffery orbits the director never crosses the shear-gradient plane). The dynamics of \textbf{n} are much more complicated. The trajectory reaches a quasi-steady state position on either hemisphere alternately every time $p_z$ changes its sign, as observed in FIG.~\ref{fig:unbounded_5_control_n_top}, and the quasi-steady state position gets closer and closer to zero, which is consistent with the damped oscillation of $n_z$ (FIG.~\ref{fig:unbounded_5_control_pz_nz_t}). All these behaviors indicate that under small perturbation, the orbit of a slightly deformable prolate spheroidal capsule evolves back towards the symmetric tumbling motion in the $x$-$y$ plane, indicating that the symmetric tumbling orbit in the $x$-$y$ plane is stable.

By further increasing the initial angle $\alpha$ of the prolate spheroid off the $x$ axis, we observe that for an arbitrary $\alpha$, the orbit of the prolate spheroid evolves towards the symmetric tumbling in the $x$-$y$ plane regardless of the initial orientation. This shows that the in-plane symmetric tumbling is the globally stable state for the prolate spheroidal capsules in the parameter regime considered. 

   \subsection{Dynamics of curved prolate spheroids}
\begin{figure}[h]
\centering
\captionsetup{justification=raggedright}
 \subfloat[]
{
    \includegraphics[width=0.40\textwidth]{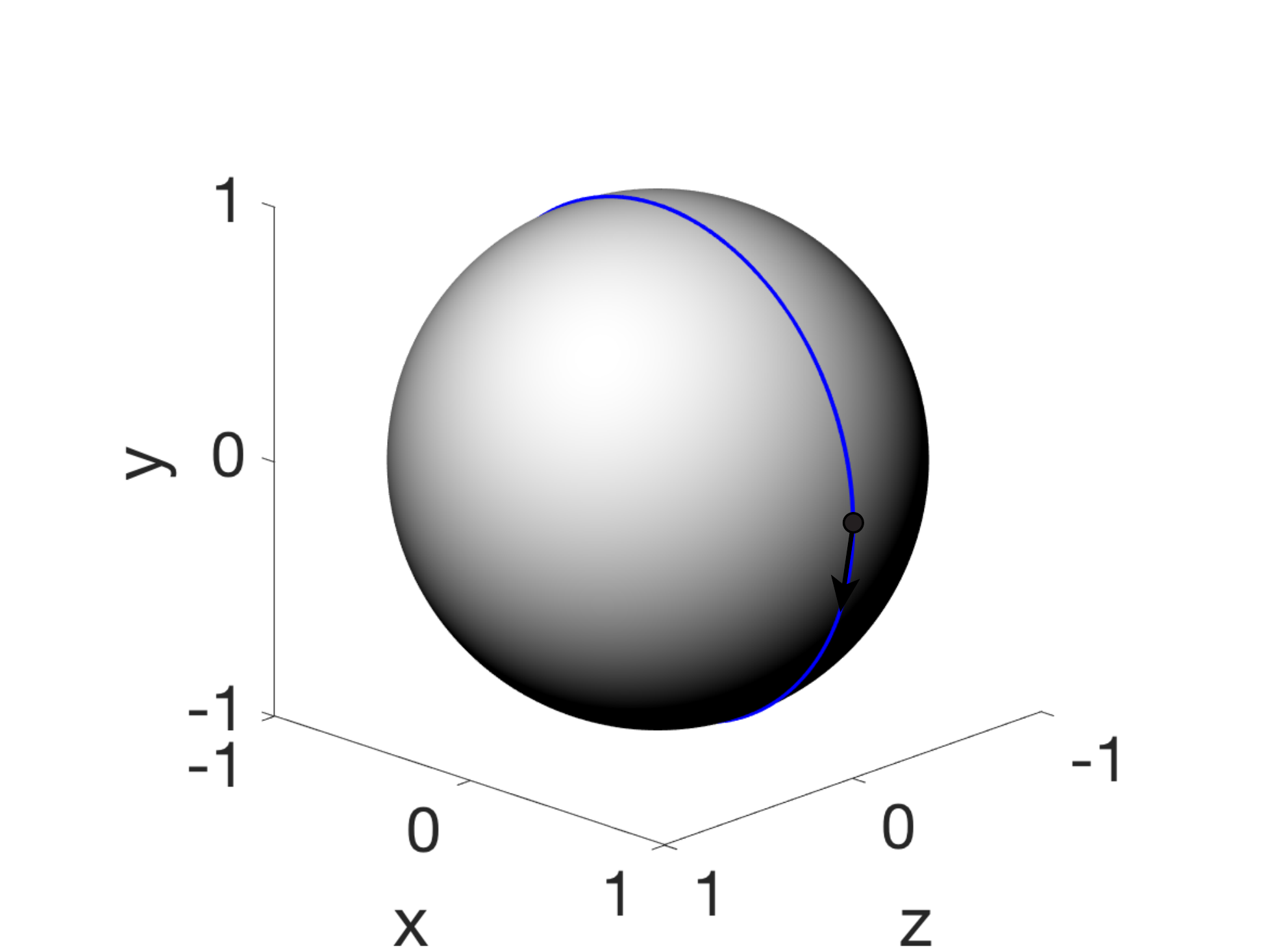}
    \label{fig:0_0_p_early}
}
\subfloat[] 
{
    \includegraphics[width=0.40\textwidth]{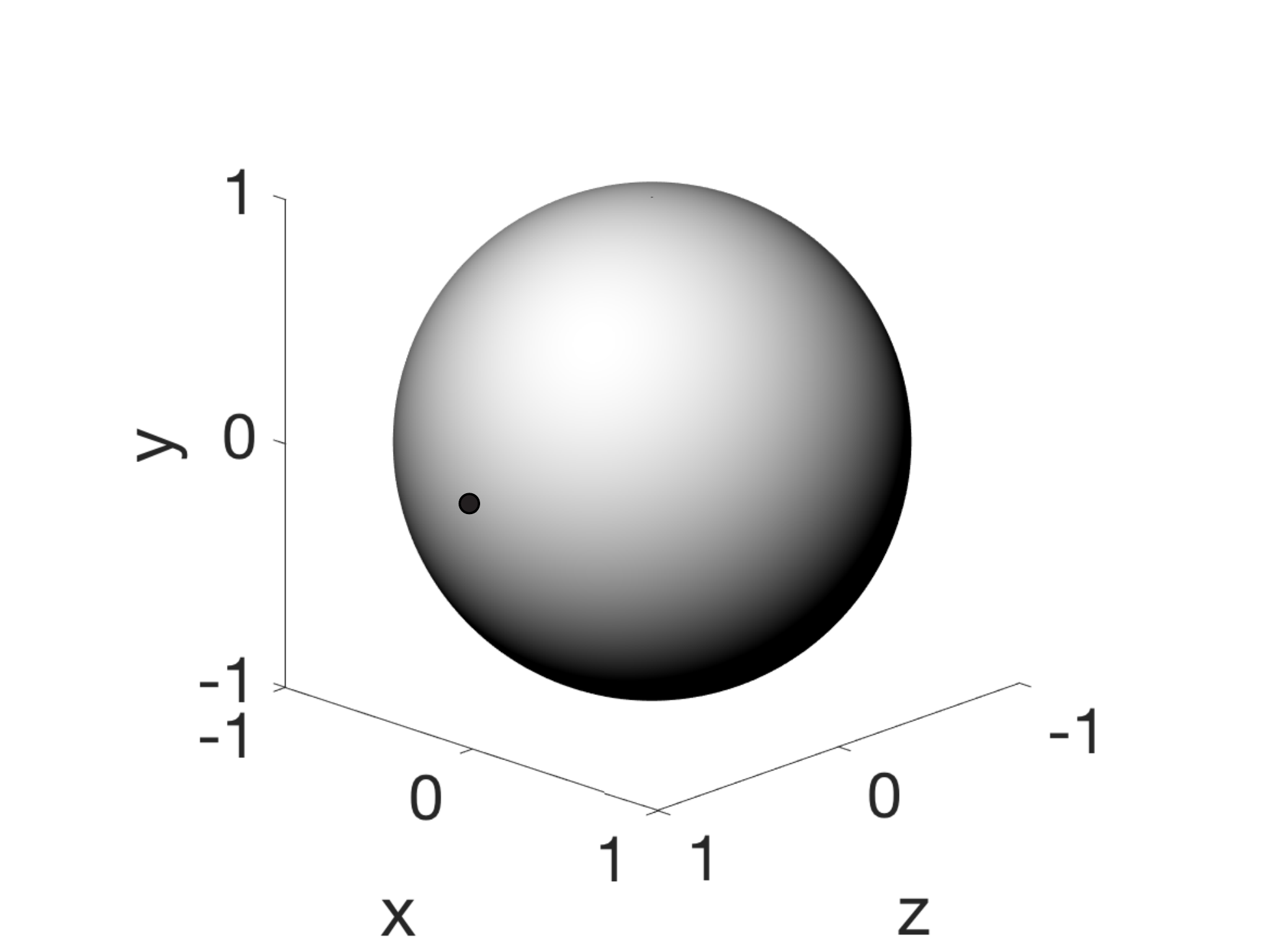}
    \label{fig:0_0_n_early}
}
\caption[0_0_trajectories]{Trajectories of \textbf{p} (a) and \textbf{n} (b) for a curved prolate capsule with initial orientation [$\alpha, \beta$] = [0,0] at the early stage of simulation ($0 < \dot{\gamma}t < 500$).}
\label{0_0_trajectories}
\end{figure}

\begin{figure}[ht]
\centering
\captionsetup{justification=raggedright}
 \subfloat[]
{
    \includegraphics[width=0.30\textwidth]{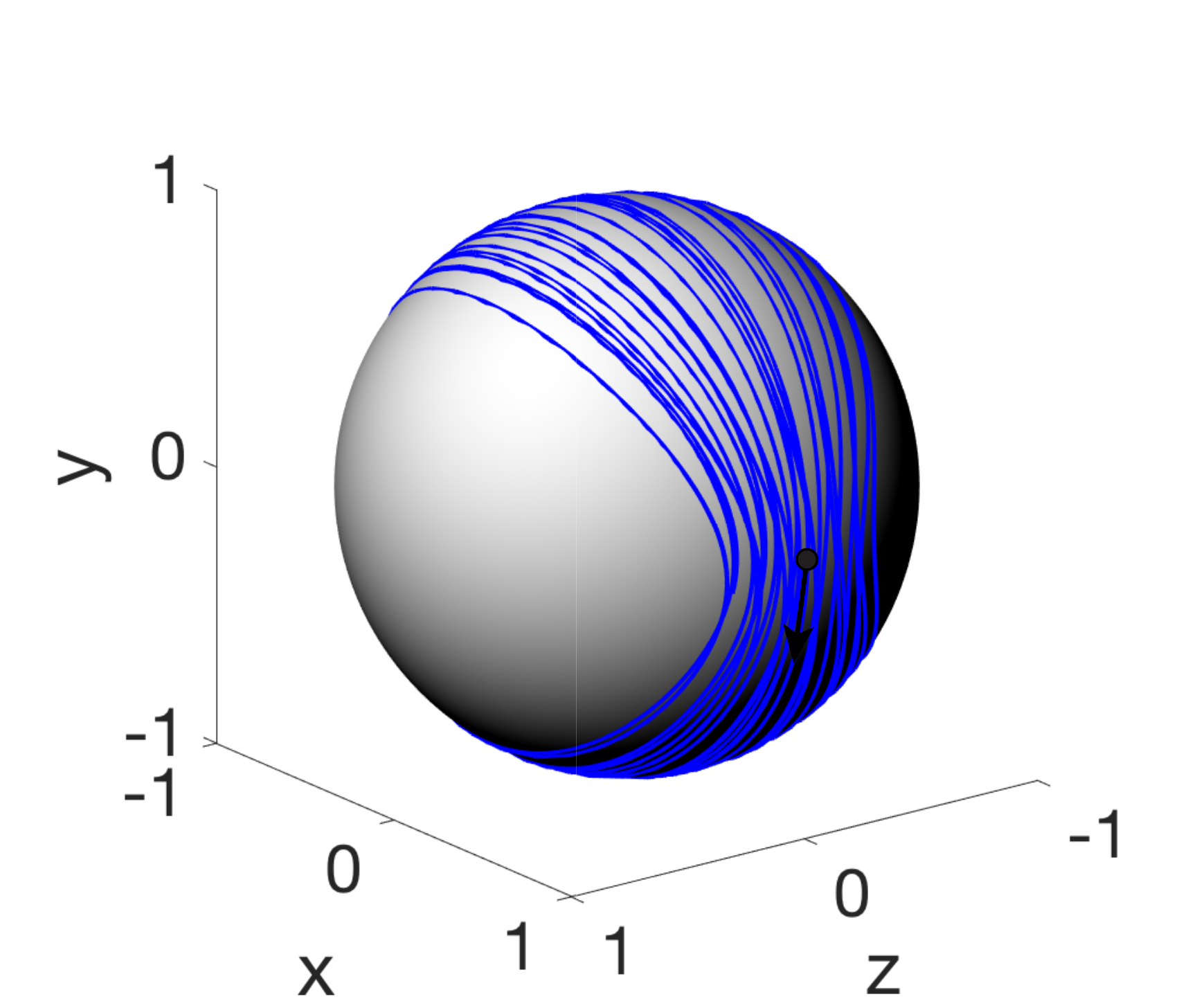}
    \label{fig:0_30_p_early}
}
 \subfloat[]
{
    \includegraphics[width=0.30\textwidth]{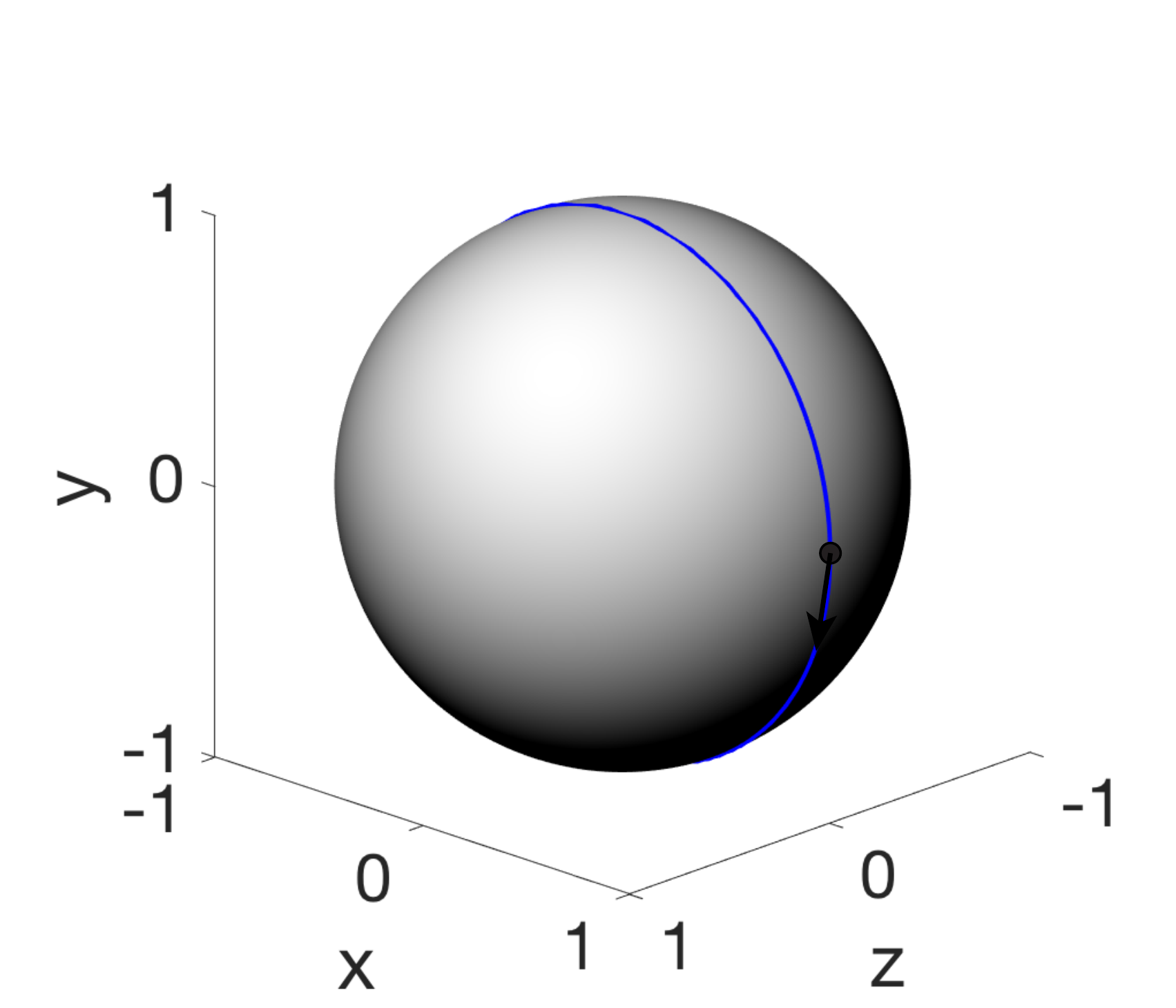}
    \label{fig:0_90_p_early}
}
 \subfloat[]
{
    \includegraphics[width=0.30\textwidth]{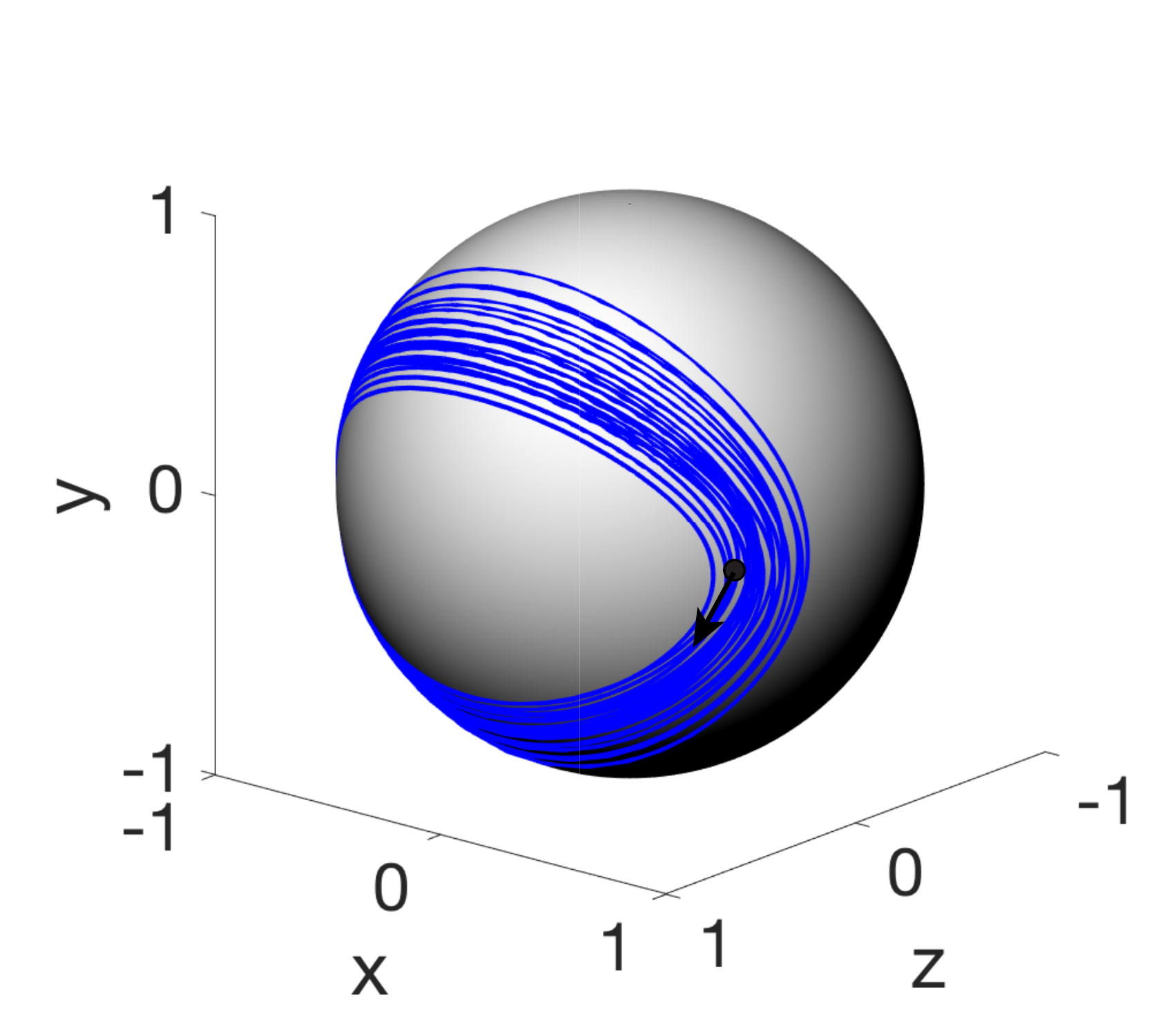}
    \label{fig:30_60_p_early}
}
\\
 \subfloat[]
{
    \includegraphics[width=0.30\textwidth]{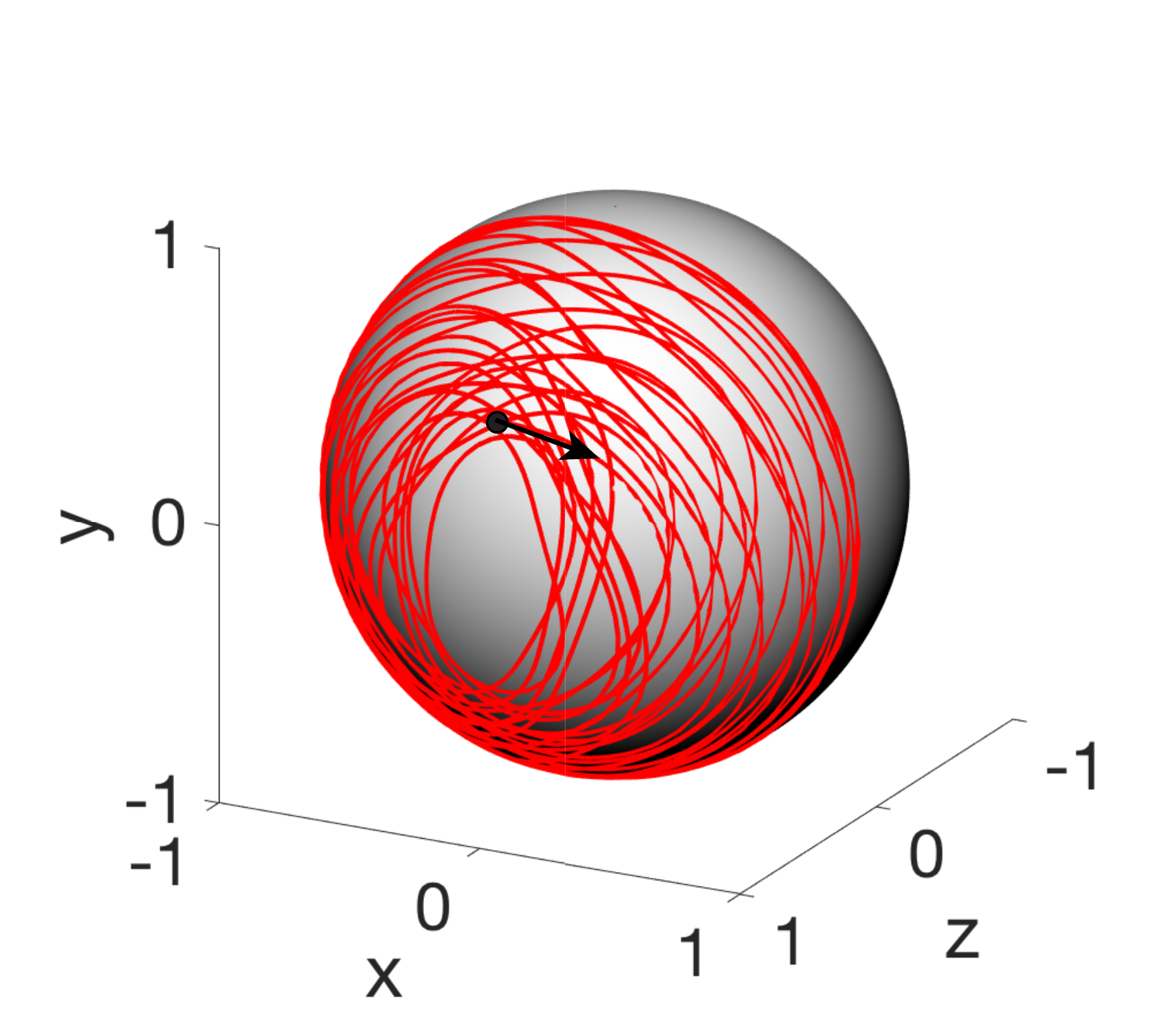}
    \label{fig:0_30_n_early}
}
 \subfloat[]
{
    \includegraphics[width=0.30\textwidth]{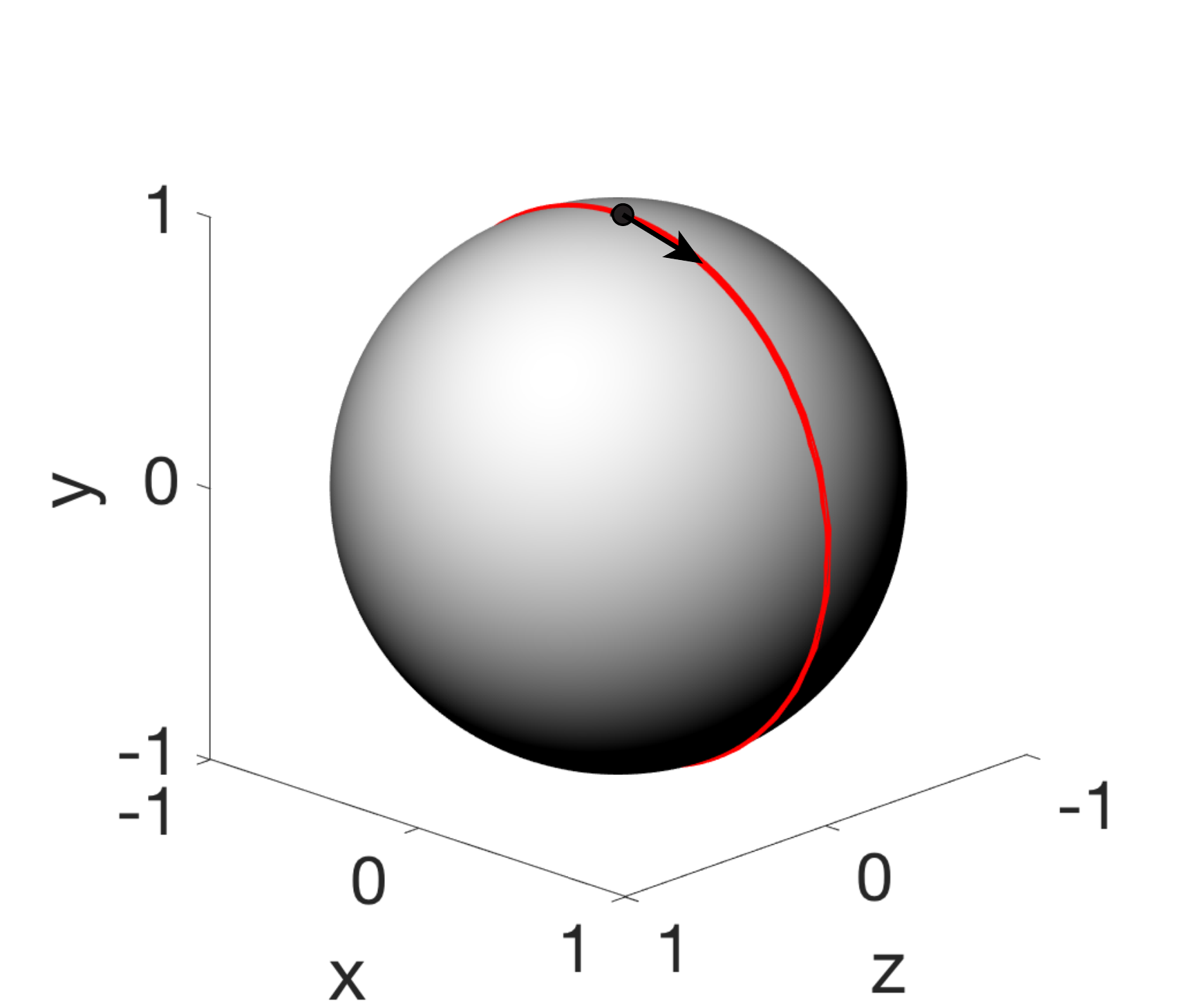}
    \label{fig:0_90_n_early}
}
 \subfloat[]
{
    \includegraphics[width=0.30\textwidth]{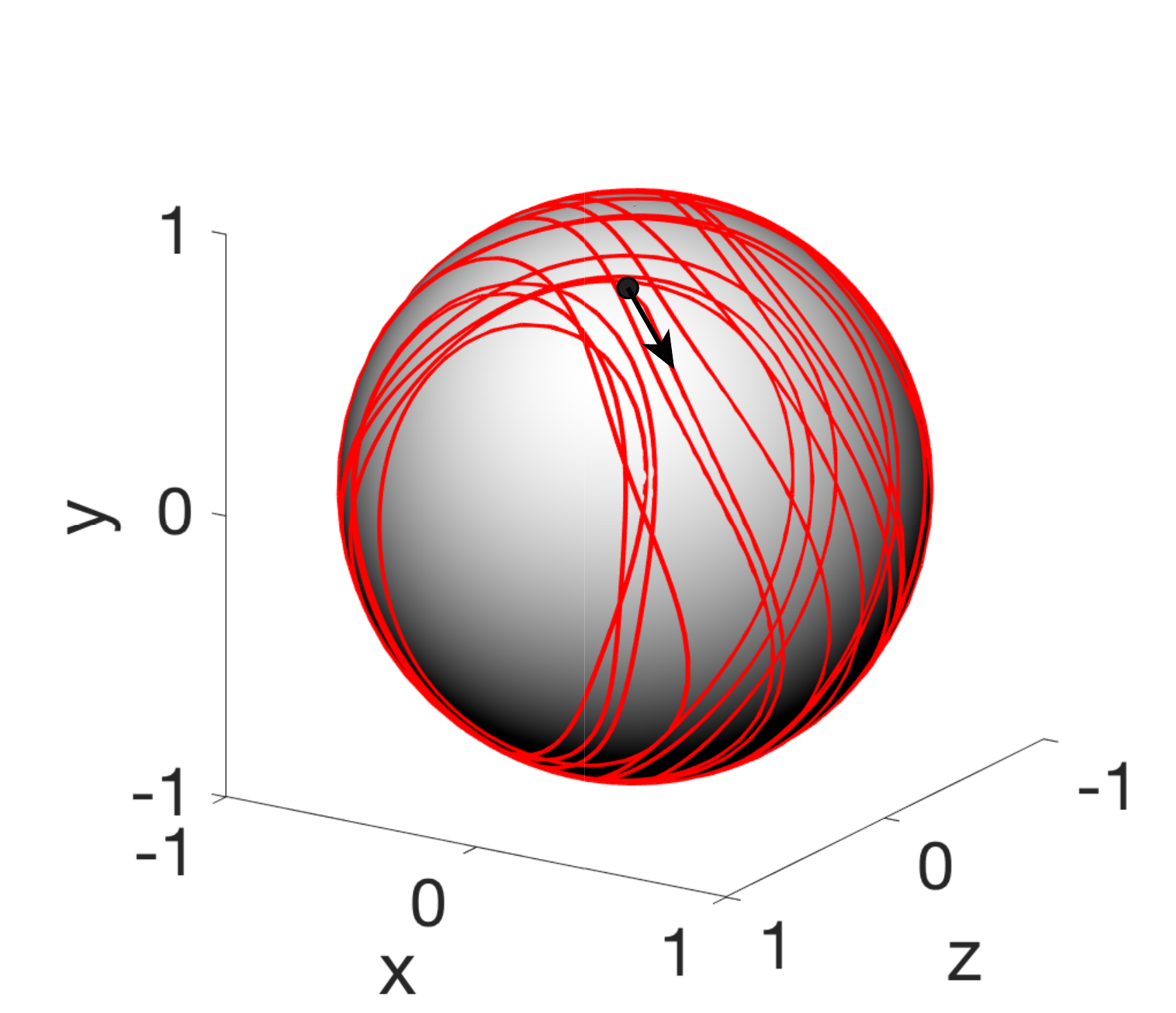}
    \label{fig:30_60_n_early}
}

\caption[early_stage_trajectories]{Early-stage trajectories on the unit sphere of the \textbf{p} (blue) and \textbf{n} (red) vectors of a curved prolate capsule with initial orientation [$\alpha, \beta$] = [0,$\pi/6$] ($0 < \dot{\gamma}t < 500$) (a,d), [0,$\pi/2$] ($0 < \dot{\gamma}t < 100$) (b,e), and [$\pi/6$,$\pi/3$] ($0 < \dot{\gamma}t < 400$) (c,f), respectively. }
\label{30_60_trajectories}
\end{figure}

\begin{figure}[h]
\centering
\captionsetup{justification=raggedright}
\subfloat[]
{
    \includegraphics[width=0.95\textwidth]{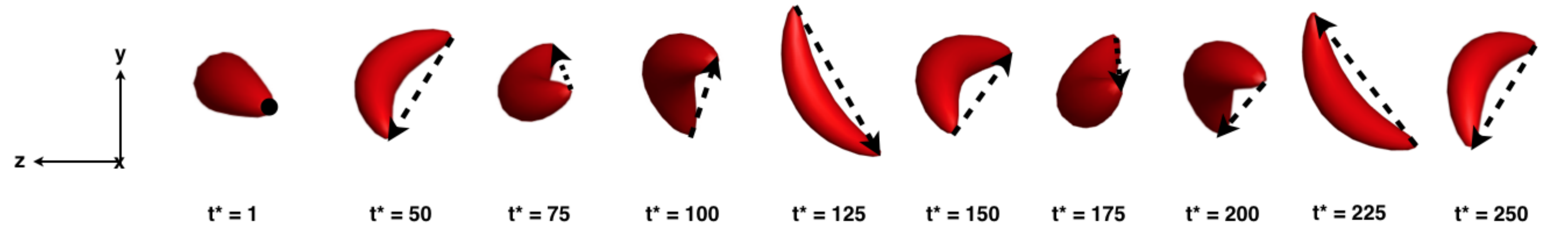}
    \label{fig:0_30_early_movie}
}
\\
\subfloat[]
{
    \includegraphics[width=0.95\textwidth]{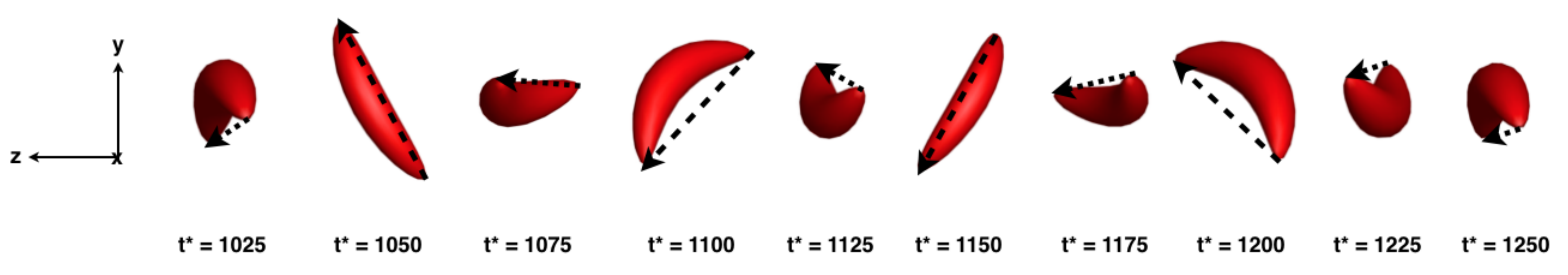}
    \label{fig:0_30_ss_movie}
}
\caption[0_30_movies]{Time sequence images (front view) of the early-stage (a) and long-time (b) motions of a curved prolate  capsule with initial orientation [$\alpha, \beta$] = [0,$\pi/6$]. The arrow in each image represents the \textbf{p} vector at the corresponding time spot.}
\label{0_30_movie}
\end{figure}

In this section, we perform a systematic investigation of the dynamics of curved prolate spheroids at low Ca as described in Section~\ref{sec:sickle_model}. We first examine the early-stage dynamics ($0 < \dot{\gamma}t \lesssim 500$) of curved prolate  capsules in the case of Ca = 0.12 and $K = 0.36$ with various initial orientations, beginning with the case where \textbf{p} is initially aligned with the $x$ direction and \textbf{n} with the $z$ direction. Here the trajectory of \textbf{p} traces out a closed circle on the unit sphere (FIG. \ref{fig:0_0_p_early}), while \textbf{n} remains fixed (FIG. \ref{fig:0_0_n_early}), which indicates that the capsule is taking a sideways Jeffery-like tumbling motion. In the case of [$\alpha, \beta$] = [0,$\pi/6$] (FIG. \ref{fig:0_30_p_early} and \ref{fig:0_30_n_early}),  \textbf{p}  is observed to cross the $x$-$y$ plane back and forth, exhibiting a wobbling motion which is not Jeffery-like, as shown in FIG. \ref{fig:0_30_early_movie}. We then consider the case where [$\alpha, \beta$] = [0,$\pi/2$], $i.e.$, the capsule is completely lying in the $x$-$y$ plane initially, with  \textbf{p}  aligned with the $x$ direction and  \textbf{n}  with the $y$ direction. The closed circular orbits on the unit sphere indicate that the capsule is taking a rigid-body-like symmetric tumbling motion as its early-stage dynamics (FIG. \ref{fig:0_90_p_early} and \ref{fig:0_90_n_early}). In the case [$\alpha, \beta$] = [$\pi/6$,$\pi/3$], a more Jeffery-like orbit is observed (FIG. \ref{fig:30_60_p_early} and \ref{fig:30_60_n_early}), in the sense that $\textbf{p}$ does not cross the shear plane. Here it is noteworthy that in a Jeffery-like orbit, the trajectory of \textbf{n} spans both hemispheres divided by the shear plane on the unit sphere (FIG. \ref{fig:30_60_n_early}), while it only appears on one hemisphere of the unit sphere in the non-Jeffery-like orbit observed in the case of [$\alpha, \beta$] = [0,$\pi/6$] described above (FIG. \ref{fig:0_30_n_early}). 

\begin{figure}[h]
\centering
\captionsetup{justification=raggedright}
 \subfloat[]
{
    \includegraphics[width=0.33\textwidth]{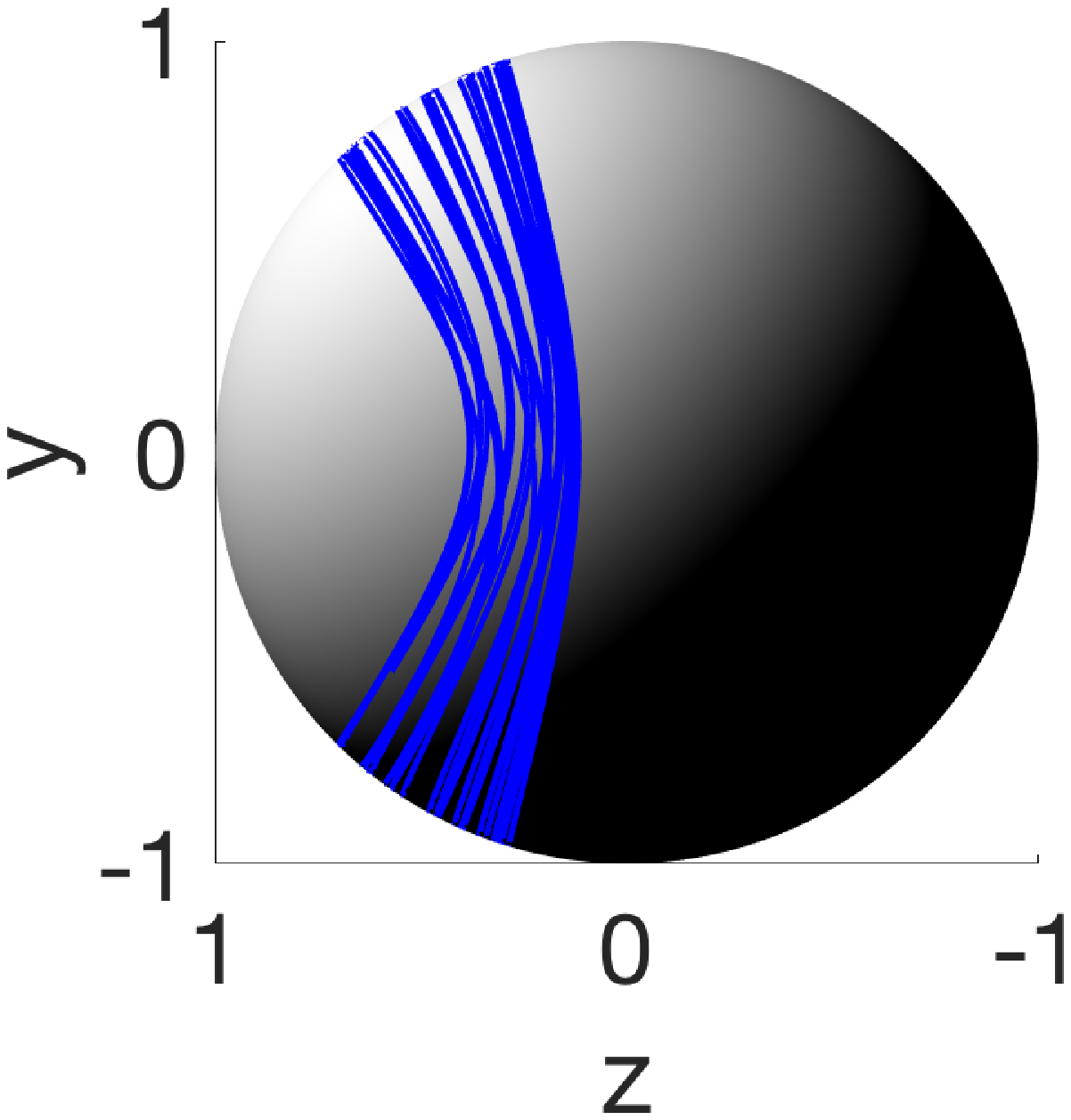}
    \label{fig:unbounded_0_30_ss}
}
\subfloat[] 
{
    \includegraphics[width=0.33\textwidth]{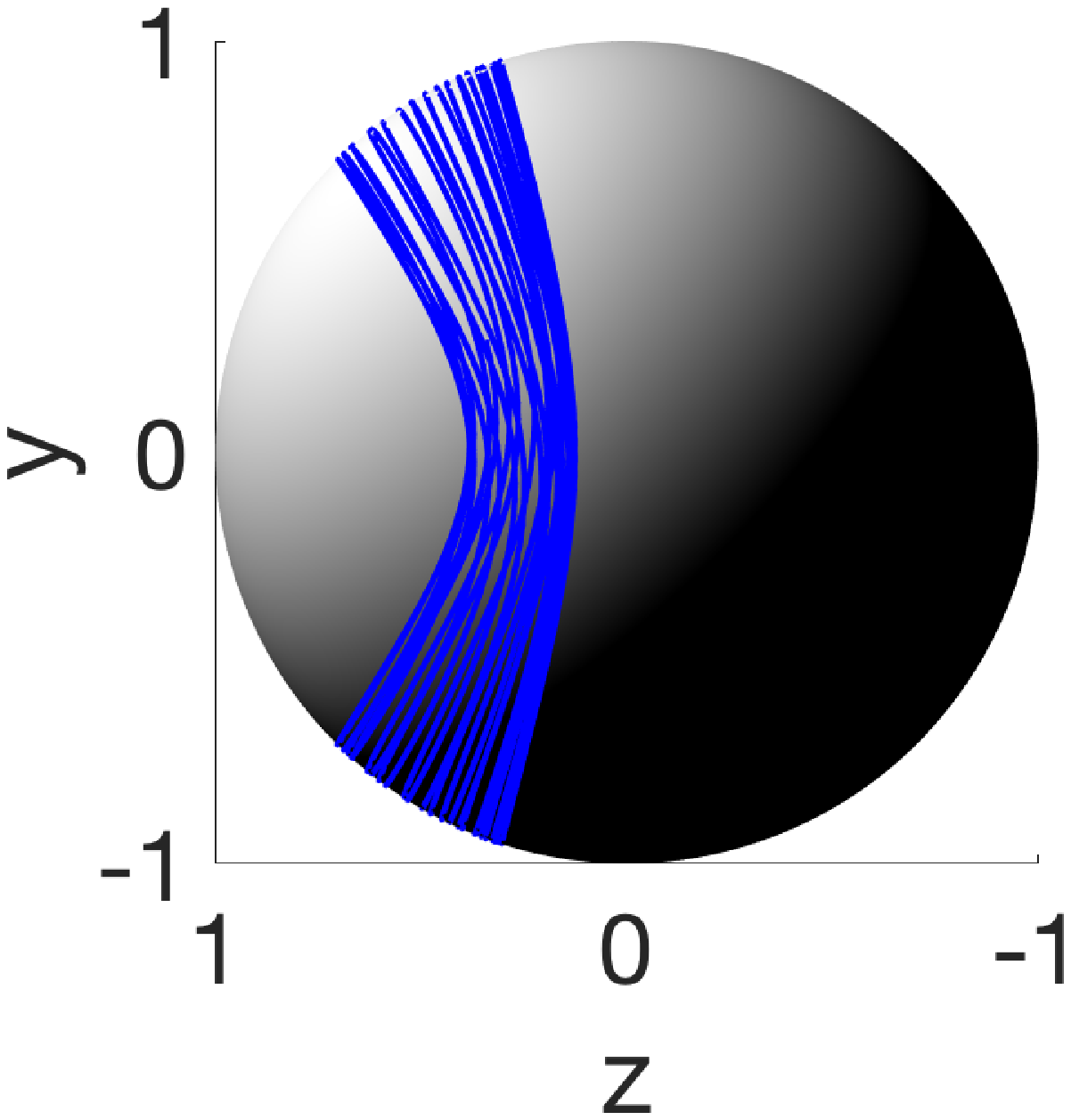}
    \label{fig:unbounded_0_90_ss}
}
\subfloat[] 
{
    \includegraphics[width=0.33\textwidth]{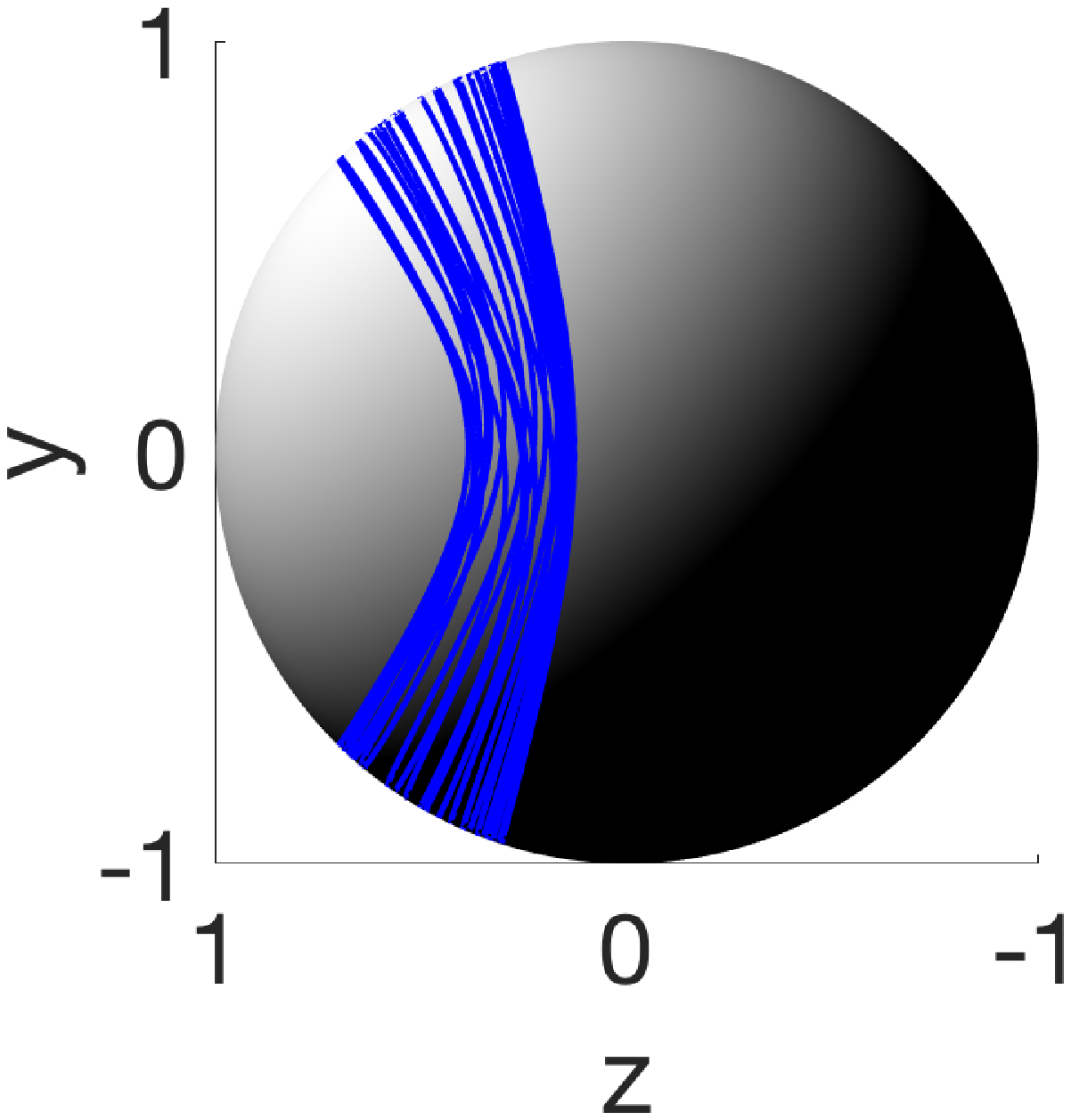}
    \label{fig:unbounded_30_60_ss}
}
\caption[unbounded_steady_state_comparison]{Long-time trajectories of \textbf{p} of curved prolate  capsules with initial orientations [$\alpha, \beta$] = [0,$\pi/6$] (a), [0,$\pi/2$] (b), and [$\pi/6$,$\pi/3$] (c) on the unit sphere ($y$-$z$ view).}
\label{unbounded_ss_p_comparison}
\end{figure}

The early-stage dynamics of the curved prolate  capsules described above, however, are all found to be long-lived transients. FIG.~\ref{unbounded_ss_p_comparison} shows the long-time ($\dot{\gamma}t > 800$) trajectories of the \textbf{p} vector of curved prolate  capsules with different initial orientations on the unit sphere. The long-time dynamics of the capsules, regardless of their early-stage orbits, all evolve into a quasi-periodic kayaking motion that might be described as a ``modulated Jeffery orbit". This long-time motion is independent of the initial orientation of the capsules, as indicated by the nearly identical trajectories of \textbf{p} on the unit sphere (FIG.~\ref{unbounded_ss_p_comparison}). Qualitatively, these orbits have a fast time scale corresponding to one cycle of kayaking, with a much longer time scale describing a slow modulation of the orbit. Each kayaking cycle is slightly different than the previous, a fact that can be attributed to the dynamics of $\textbf{n}$, which are coupled to those of $\textbf{p}$ for a curved particle. FIG.~\ref{fig:0_30_ss_movie} shows snapshots of a long-time orbit.  

\begin{figure}[t]
\centering
\captionsetup{justification=raggedright}
 \subfloat[]
{
    \includegraphics[width=0.6\textwidth]{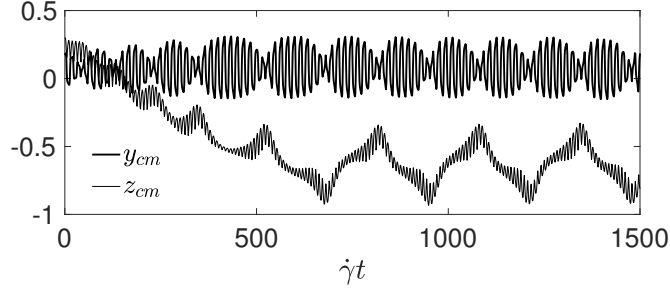}
    \label{fig:unbounded_0_30_ss_cm}
}
\\
\subfloat[] 
{
    \includegraphics[width=0.6\textwidth]{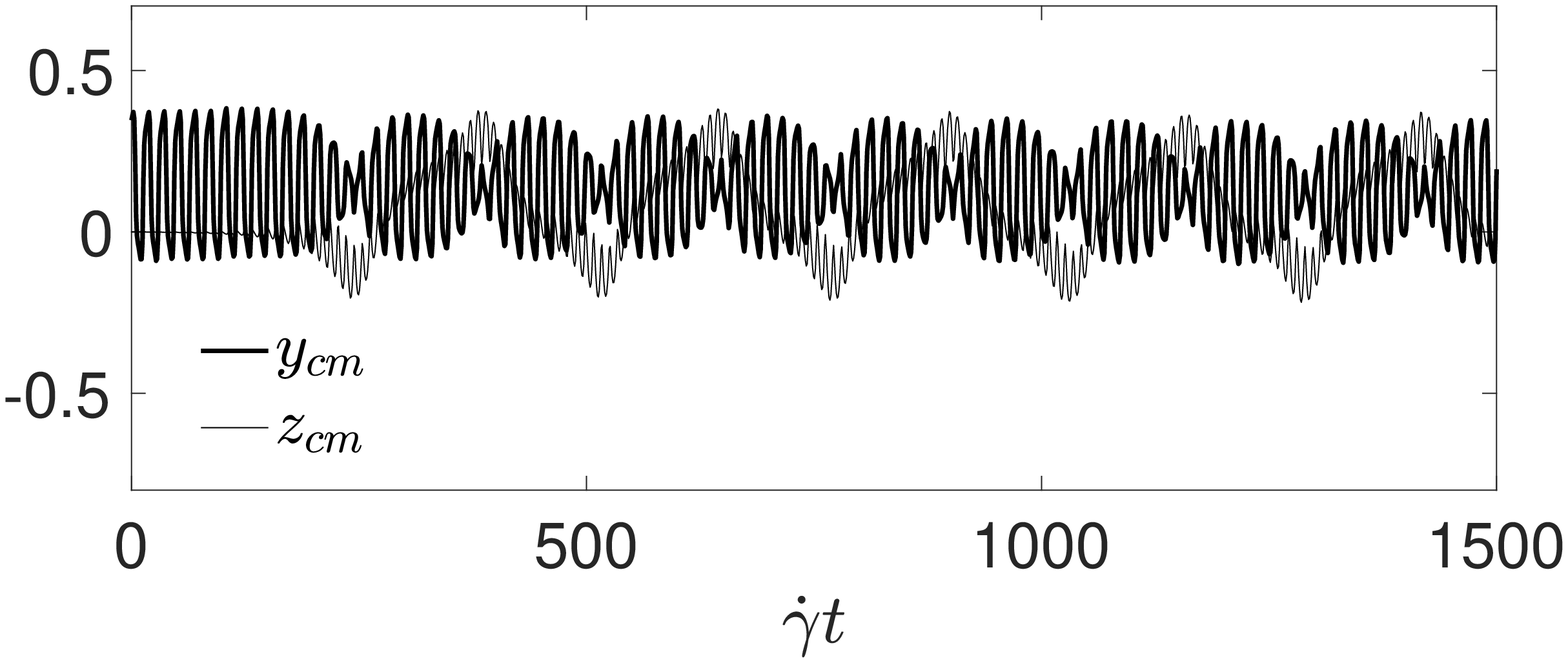}
    \label{fig:unbounded_0_90_ss_cm}
}
\\
\subfloat[] 
{
    \includegraphics[width=0.6\textwidth]{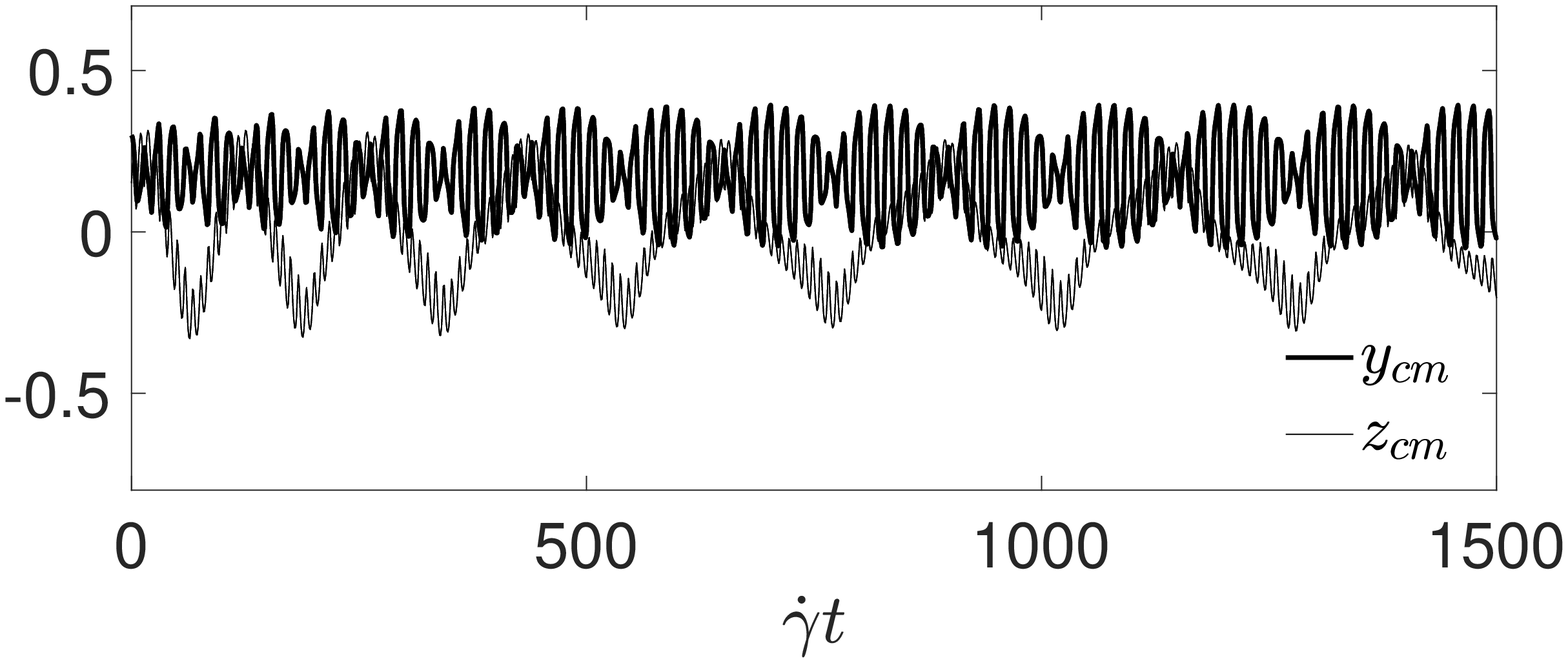}
    \label{fig:unbounded_30_60_ss_cm}
}
  \caption[center-of-mass positions at steady state]{Evolution of $y$- and $z$-components $y_{cm}$ and $z_{cm}$ of the center-of-mass position of curved prolate capsules with initial orientations [$\alpha, \beta$] = [0,$\pi/6$] (a), [0,$\pi/2$] (b), and [$\pi/6$,$\pi/3$] (c).}
  \label{unbounded_ss_cm}
  \end{figure}

The two time scales noted above are also observed in the long-time center-of-mass position profiles shown in FIG.~\ref{unbounded_ss_cm}, with a shorter time scale representing one cycle of the kayaking motion described above superimposed onto a longer time scale corresponding to the orbit modulation. In the parameter regime considered here, we have seen no evidence of drift in the $y$ (``wall normal'') direction. This is in contrast to the observations of \cite{Wang:2012ji} for curved rigid fibers in the ``flipping and scooping'' regime; we believe the difference arises from the modest aspect ratios considered here, which are much smaller than those studied by \cite{Wang:2012ji}. We do, however, observe a transient phase of spanwise drift in the case of the capsule initialized to perform a sideways tumbling motion. This vanishes once the cell reaches its long-time orientational trajectory. Although there is no drift at long time, both the $y$ and $z$ components of the center of mass oscillate as the cell kayaks.  

\begin{figure}[h]
\centering
\captionsetup{justification=raggedright}
\includegraphics[width=0.65\textwidth]{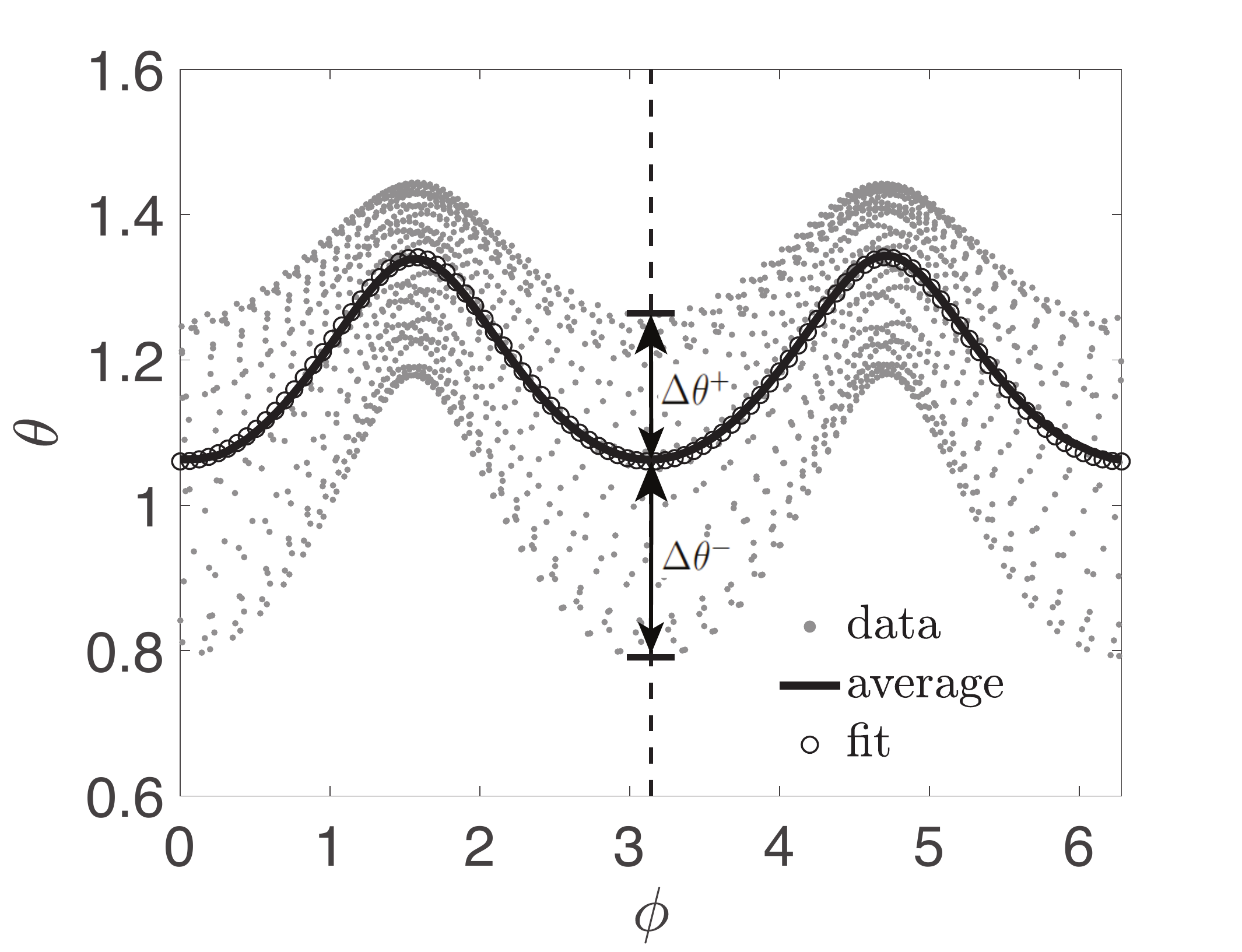}
\caption[theta_phi]{The long-time trajectory (grey dots) of a curved prolate  capsule (Ca = 0.12, $K$ = 0.36) in terms of $\phi$ and $\theta$ and the average trajectory (black line). Black circles represent the fit of the average trajectory using Eq.~\ref{eq:Jeffery_orbit_2}.}
\label{fig:theta_phi_Ca_0_12_K_0_36}
\end{figure}

\begin{figure}[h]
\centering
\captionsetup{justification=raggedright}
 \subfloat[]
{
    \includegraphics[width=0.45\textwidth]{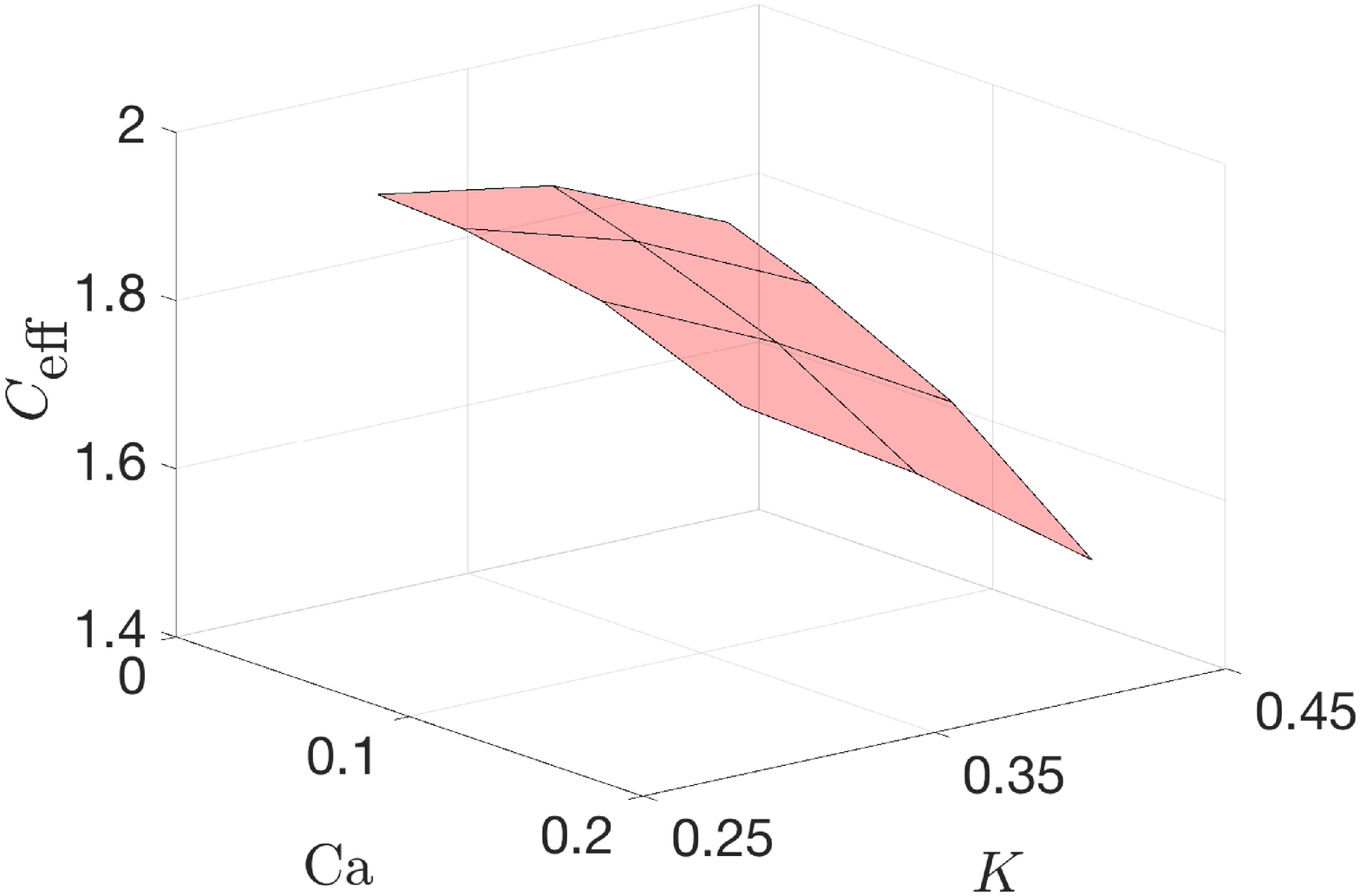}
    \label{fig:C_dependencies}
}
\subfloat[] 
{
    \includegraphics[width=0.42\textwidth]{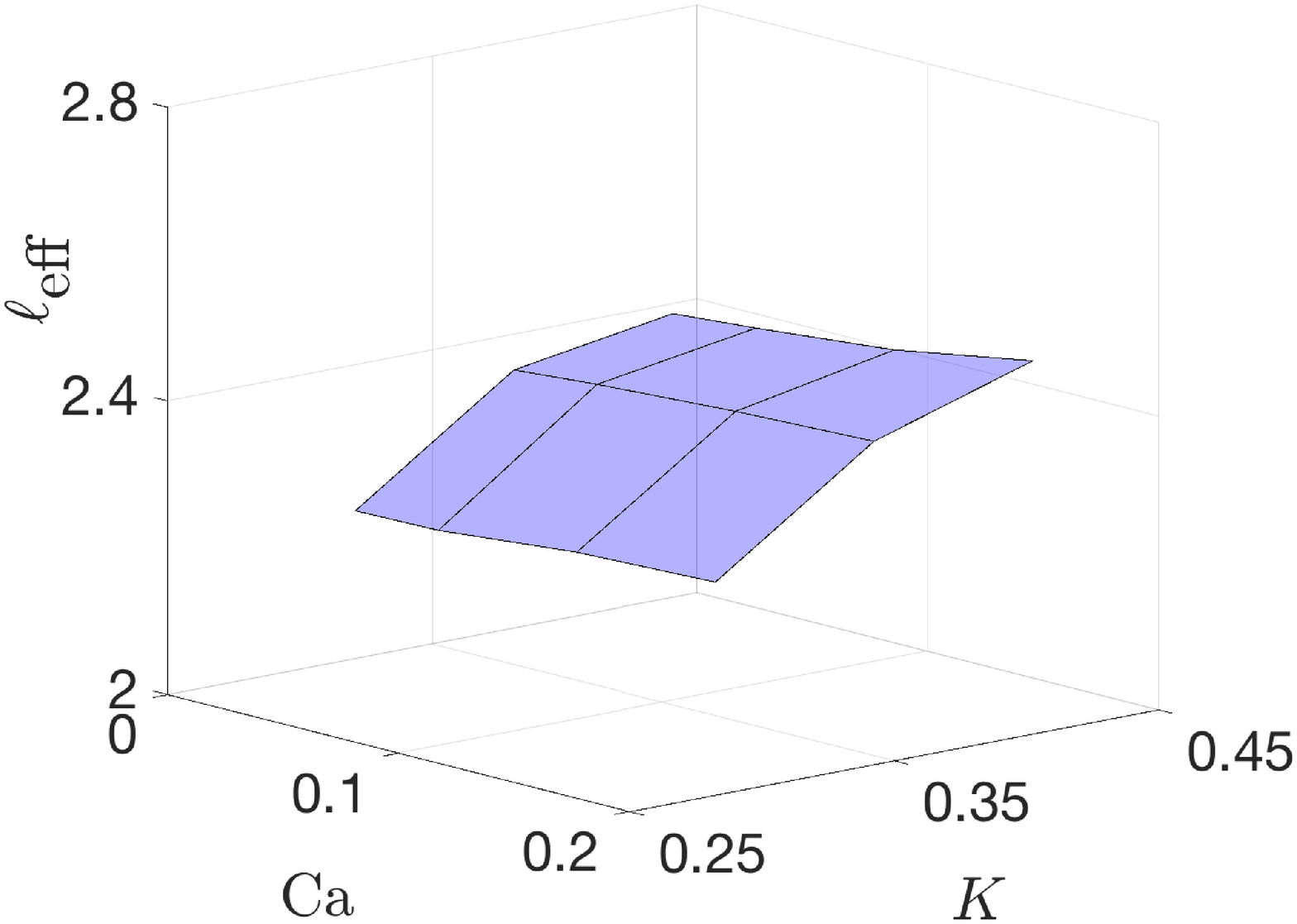}
    \label{fig:l_dependencies}
}
\caption[dependencies_on_Ca_and_K]{The dependencies of (a) the effective orbit constant $C_{\textnormal{eff}}$ of the long-time orbit and (b) the effective aspect ratio $\ell_{\textnormal{eff}}$ on the deformability Ca and the degree of curvature $K$ of the curved prolate capsules.}
\label{fig:dependencies_on_Ca_and_K}
\end{figure}
  
Now we vary $\Ca$ and $K$ to see the effects of deformability and curvature on the long-time dynamics. To illustrate how we characterize these effects, we first revisit the case of $\Ca = 0.12$, $K = 0.36$, for which the orbits in FIG.~\ref{unbounded_ss_p_comparison} show the trajectory of $\textbf{p}$ on the unit sphere. The grey dotted curve in FIG.~\ref{fig:theta_phi_Ca_0_12_K_0_36} shows this orbit expressed in spherical coordinates $\phi$ and $\theta$ based on the coordinate system used in Jeffery's theory \cite{Jeffery:1922wb}; here $\phi$ is the azimuthal angle with respect to the $y$ axis and $\theta$ is the polar angle with respect to the $z$ axis. With this representation, we can find a mean trajectory $\bar \theta (\phi)$  by averaging $\theta$ at each $\phi$. This is the black curve in FIG.~\ref{fig:theta_phi_Ca_0_12_K_0_36}. For a true Jeffery orbit of a rigid particle, the relation between the two angles $\phi$ and $\theta$ is given by 
\begin{equation} \label{eq:Jeffery_orbit_2}
  \mathrm{tan} \; \theta = \frac{C \ell}{\big(\ell^2 \; \mathrm{cos}^2 \phi + \mathrm{sin}^2 \phi \big)^{1/2}},
\end{equation}
where $C$ and $\ell$ are the orbit constant and aspect ratio for the rigid spheroid. We have found that in all cases considered here, we can approximate with good accuracy the curve $\bar \theta (\phi)$ as a Jeffery orbit for a spheroid with an effective orbit constant $C_{\textnormal{eff}}$ and effective aspect ratio $\ell_{\textnormal{eff}}$. That is, we can use  $C_{\textnormal{eff}}$ and $\ell_{\textnormal{eff}}$ as fitting parameters in the expression 
\newcommand{\eff}{\mathrm{eff}}
\begin{equation} \label{eq:Jeffery_orbit_2}
  \mathrm{tan} \; \bar{\theta} = \frac{C_\eff \ell_\eff}{\big(\ell_{\eff}^2 \; \mathrm{cos}^2 \phi + \mathrm{sin}^2 \phi \big)^{1/2}}.
\end{equation}
The black circles in FIG.~\ref{fig:theta_phi_Ca_0_12_K_0_36} show this approximation for $\Ca = 0.12$, $K = 0.36$. 

We have computed long-time trajectories over a range of $\Ca$ and $K$ and used the resulting values of $C_\eff$ and $\ell_\eff$ to characterize the dynamics. These results are  summarized in FIG.~\ref{fig:dependencies_on_Ca_and_K}. It is observed that $\ell_{\textnormal{eff}}$ increases weakly as $K$ increases but shows very little dependency on $\Ca$ (FIG.~\ref{fig:l_dependencies}). More importantly, $C_{\textnormal{eff}}$ decreases as either $\Ca$ or $K$ increases (FIG.~\ref{fig:C_dependencies}), indicating a shift of the orbit towards log-rolling motion with increasing curvature or deformability. 

For completeness, we briefly consider the deviations of the instantaneous trajectories from the mean curve $\bar{\theta}(\phi)$. The maximum positive and negative deviations from the mean when $\phi=\pi$ are denoted as $\Delta \theta^+$ and $\Delta \theta^-$. The relative deviations $\Delta \theta / \overline{\theta}$ are found to be insensitive to the parameters in the range considered here, with $ 0.16 < \Delta \theta^+ / \overline{\theta} < 0.19$, and $-0.35 < \Delta \theta^- / \overline{\theta} < -0.30$. 
      
\section{CONCLUSION} \label{sec:conclusion}
We investigated the dynamics of neutrally-buoyant, slightly deformable straight and curved prolate  capsules subjected to unbounded simple shear flow at zero Reynolds number using direct simulations. The curved prolate capsules in this work serve as a model for the typical crescent-shaped sickle RBCs in sickle cell disease (SCD), and we aimed to show the effects of initial orientation, membrane deformability and curvature of the capsules on their dynamics.

The dynamics of the capsules are simulated using a full model incorporating both shear and bending elasticities. The parameter regime is based on the range of experimentally determined values for blood flow in the microcirculation. We revealed that with low deformability, straight prolate spheroidal capsules take tumbling in the shear plane as the unique globally stable orbit independent of the initial orientation. As the trajectories of the straight prolate capsules transiently evolve towards this in-plane tumbling, the director can cross the $x$-$y$ plane back and forth, showing a non-Jeffery-like motion.

Curved prolate capsules exhibit complicated and interesting dynamics due to the combined effects of membrane deformability $\Ca$ and the curvature $K$ of the capsules. At short times, a non-Jeffery-like behavior, in which the director (end-to-end vector) of the capsule crosses the shear-gradient plane periodically, is observed for specific initial orientations. Steady drift of the center of mass in the vorticity direction is also observed for some initial orientations. At long times, however, the trajectories of the curved capsules, regardless of their initial orientations, all evolve towards a Jeffery-like quasi-periodic kayaking orbit with the center-of-mass position of a curved capsule crossing the shear plane periodically instead of remaining in it. No cross-stream drift is found. The average of the long-time orbit can be well approximated using the analytical solution for Jeffery orbits, yielding an effective orbit constant $C_{\textnormal{eff}}$ and aspect ratio $\ell_{\textnormal{eff}}$. A parameter study further reveals that $C_{\textnormal{eff}}$ decreases as the capsule becomes more deformable or curved, indicating a shift of the orbit towards log-rolling motion. 

Overall, these results represent an effort to understand the motion of single sickle RBCs, taking into account the pathologically altered shape as well as the elastic and bending moduli of these diseased cells. Major aspects of the pathophysiology of SCD remain poorly understood and currently, there is an increased focus on how aberrant cell interactions, between sickle RBCs and other blood cells ($i.e.$ platelets and leukocytes) or between sickle RBCs and endothelial cells, may contribute to phenomena such as thrombosis or vascular inflammation that are known to be vital part of the process of SCD. In particular, the effective orbit constant of a sickle cell moving near a blood vessel wall may be reflected in the propensity for that cell to collide with and potentially damage the wall. More broadly, the results of this work demonstrate how the mechanical properties and distorted cell shape of sickle RBCs lead to distinct trajectories that may, in turn, cause pathologic biophysical activation of other cells, and may provide insight into how these disparate aspects of SCD are ultimately linked. That knowledge will then lead to improved methods to diagnose complications of and new therapeutic targets for SCD.

\begin{acknowledgments}
This work was supported by NSF Grant No.~CBET-1436082 and NIH Grant No.~R21MD011590-01A1.
\end{acknowledgments}


\begin{thebibliography}{36}%
\makeatletter
\providecommand \@ifxundefined [1]{%
 \@ifx{#1\undefined}
}%
\providecommand \@ifnum [1]{%
 \ifnum #1\expandafter \@firstoftwo
 \else \expandafter \@secondoftwo
 \fi
}%
\providecommand \@ifx [1]{%
 \ifx #1\expandafter \@firstoftwo
 \else \expandafter \@secondoftwo
 \fi
}%
\providecommand \natexlab [1]{#1}%
\providecommand \enquote  [1]{``#1''}%
\providecommand \bibnamefont  [1]{#1}%
\providecommand \bibfnamefont [1]{#1}%
\providecommand \citenamefont [1]{#1}%
\providecommand \href@noop [0]{\@secondoftwo}%
\providecommand \href [0]{\begingroup \@sanitize@url \@href}%
\providecommand \@href[1]{\@@startlink{#1}\@@href}%
\providecommand \@@href[1]{\endgroup#1\@@endlink}%
\providecommand \@sanitize@url [0]{\catcode `\\12\catcode `\$12\catcode
  `\&12\catcode `\#12\catcode `\^12\catcode `\_12\catcode `\%12\relax}%
\providecommand \@@startlink[1]{}%
\providecommand \@@endlink[0]{}%
\providecommand \url  [0]{\begingroup\@sanitize@url \@url }%
\providecommand \@url [1]{\endgroup\@href {#1}{\urlprefix }}%
\providecommand \urlprefix  [0]{URL }%
\providecommand \Eprint [0]{\href }%
\providecommand \doibase [0]{http://dx.doi.org/}%
\providecommand \selectlanguage [0]{\@gobble}%
\providecommand \bibinfo  [0]{\@secondoftwo}%
\providecommand \bibfield  [0]{\@secondoftwo}%
\providecommand \translation [1]{[#1]}%
\providecommand \BibitemOpen [0]{}%
\providecommand \bibitemStop [0]{}%
\providecommand \bibitemNoStop [0]{.\EOS\space}%
\providecommand \EOS [0]{\spacefactor3000\relax}%
\providecommand \BibitemShut  [1]{\csname bibitem#1\endcsname}%
\let\auto@bib@innerbib\@empty
\bibitem [{\citenamefont {Mohandas}\ and\ \citenamefont
  {Evans}(1994)}]{Evans1994}%
  \BibitemOpen
  \bibfield  {author} {\bibinfo {author} {\bibfnamefont {N.}~\bibnamefont
  {Mohandas}}\ and\ \bibinfo {author} {\bibfnamefont {E.}~\bibnamefont
  {Evans}},\ }\bibfield  {title} {\enquote {\bibinfo {title} {Mechanical
  properties of the red cell membrane in relation to molecular structure and
  genetic defects},}\ }\href@noop {} {\bibfield  {journal} {\bibinfo  {journal}
  {Annu. Rev. Biophys. Biomol. Struct.}\ }\textbf {\bibinfo {volume} {23}},\
  \bibinfo {pages} {787--818} (\bibinfo {year} {1994})}\BibitemShut {NoStop}%
\bibitem [{\citenamefont {Goldsmith}\ \emph {et~al.}(1972)\citenamefont
  {Goldsmith}, \citenamefont {Marlow},\ and\ \citenamefont
  {MacIntosh}}]{Goldsmith351}%
  \BibitemOpen
  \bibfield  {author} {\bibinfo {author} {\bibfnamefont {H.~L.}\ \bibnamefont
  {Goldsmith}}, \bibinfo {author} {\bibfnamefont {J.}~\bibnamefont {Marlow}}, \
  and\ \bibinfo {author} {\bibfnamefont {F.~C.}\ \bibnamefont {MacIntosh}},\
  }\bibfield  {title} {\enquote {\bibinfo {title} {Flow behaviour of
  erythrocytes. {I}. {R}otation and deformation in dilute suspensions},}\
  }\href@noop {} {\bibfield  {journal} {\bibinfo  {journal} {Proc. R. Soc.
  Lond., B, Biol. Sci.}\ }\textbf {\bibinfo {volume} {182}},\ \bibinfo {pages}
  {351--384} (\bibinfo {year} {1972})}\BibitemShut {NoStop}%
\bibitem [{\citenamefont {Fischer}\ \emph {et~al.}(1978)\citenamefont
  {Fischer}, \citenamefont {Stohr-Lissen},\ and\ \citenamefont
  {Schmid-Schonbein}}]{Fischer894}%
  \BibitemOpen
  \bibfield  {author} {\bibinfo {author} {\bibfnamefont {T.~M.}\ \bibnamefont
  {Fischer}}, \bibinfo {author} {\bibfnamefont {M.}~\bibnamefont
  {Stohr-Lissen}}, \ and\ \bibinfo {author} {\bibfnamefont {H.}~\bibnamefont
  {Schmid-Schonbein}},\ }\bibfield  {title} {\enquote {\bibinfo {title} {The
  red cell as a fluid droplet: tank tread-like motion of the human erythrocyte
  membrane in shear flow},}\ }\href@noop {} {\bibfield  {journal} {\bibinfo
  {journal} {Science}\ }\textbf {\bibinfo {volume} {202}},\ \bibinfo {pages}
  {894--896} (\bibinfo {year} {1978})}\BibitemShut {NoStop}%
\bibitem [{\citenamefont {Abkarian}\ \emph {et~al.}(2007)\citenamefont
  {Abkarian}, \citenamefont {Faivre},\ and\ \citenamefont
  {Viallat}}]{Viallat2007PRL}%
  \BibitemOpen
  \bibfield  {author} {\bibinfo {author} {\bibfnamefont {M.}~\bibnamefont
  {Abkarian}}, \bibinfo {author} {\bibfnamefont {M.}~\bibnamefont {Faivre}}, \
  and\ \bibinfo {author} {\bibfnamefont {A.}~\bibnamefont {Viallat}},\
  }\bibfield  {title} {\enquote {\bibinfo {title} {Swinging of red blood cells
  under shear flow},}\ }\href {\doibase 10.1103/PhysRevLett.98.188302}
  {\bibfield  {journal} {\bibinfo  {journal} {Phys. Rev. Lett.}\ }\textbf
  {\bibinfo {volume} {98}},\ \bibinfo {pages} {188302} (\bibinfo {year}
  {2007})}\BibitemShut {NoStop}%
\bibitem [{\citenamefont {Skotheim}\ and\ \citenamefont
  {Secomb}(2007)}]{Secomb2007PRL}%
  \BibitemOpen
  \bibfield  {author} {\bibinfo {author} {\bibfnamefont {J.~M.}\ \bibnamefont
  {Skotheim}}\ and\ \bibinfo {author} {\bibfnamefont {T.~W.}\ \bibnamefont
  {Secomb}},\ }\bibfield  {title} {\enquote {\bibinfo {title} {Red blood cells
  and other nonspherical capsules in shear flow: Oscillatory dynamics and the
  tank-treading-to-tumbling transition},}\ }\href {\doibase
  10.1103/PhysRevLett.98.078301} {\bibfield  {journal} {\bibinfo  {journal}
  {Phys. Rev. Lett.}\ }\textbf {\bibinfo {volume} {98}},\ \bibinfo {pages}
  {078301} (\bibinfo {year} {2007})}\BibitemShut {NoStop}%
\bibitem [{\citenamefont {Dupire}\ \emph {et~al.}(2010)\citenamefont {Dupire},
  \citenamefont {Abkarian},\ and\ \citenamefont {Viallat}}]{Viallat2010PRL}%
  \BibitemOpen
  \bibfield  {author} {\bibinfo {author} {\bibfnamefont {J.}~\bibnamefont
  {Dupire}}, \bibinfo {author} {\bibfnamefont {M.}~\bibnamefont {Abkarian}}, \
  and\ \bibinfo {author} {\bibfnamefont {A.}~\bibnamefont {Viallat}},\
  }\bibfield  {title} {\enquote {\bibinfo {title} {Chaotic dynamics of red
  blood cells in a sinusoidal flow},}\ }\href {\doibase
  10.1103/PhysRevLett.104.168101} {\bibfield  {journal} {\bibinfo  {journal}
  {Phys. Rev. Lett.}\ }\textbf {\bibinfo {volume} {104}},\ \bibinfo {pages}
  {168101} (\bibinfo {year} {2010})}\BibitemShut {NoStop}%
\bibitem [{\citenamefont {Fedosov}\ \emph {et~al.}(2010)\citenamefont
  {Fedosov}, \citenamefont {Caswell},\ and\ \citenamefont
  {Karniadakis}}]{FEDOSOV20102215}%
  \BibitemOpen
  \bibfield  {author} {\bibinfo {author} {\bibfnamefont {D.~A.}\ \bibnamefont
  {Fedosov}}, \bibinfo {author} {\bibfnamefont {B.}~\bibnamefont {Caswell}}, \
  and\ \bibinfo {author} {\bibfnamefont {G.~E.}\ \bibnamefont {Karniadakis}},\
  }\bibfield  {title} {\enquote {\bibinfo {title} {A multiscale red blood cell
  model with accurate mechanics, rheology, and dynamics},}\ }\href {\doibase
  https://doi.org/10.1016/j.bpj.2010.02.002} {\bibfield  {journal} {\bibinfo
  {journal} {Biophys. J.}\ }\textbf {\bibinfo {volume} {98}},\ \bibinfo {pages}
  {2215--2225} (\bibinfo {year} {2010})}\BibitemShut {NoStop}%
\bibitem [{\citenamefont {Fedosov}\ \emph {et~al.}(2011)\citenamefont
  {Fedosov}, \citenamefont {Lei}, \citenamefont {Caswell}, \citenamefont
  {Suresh},\ and\ \citenamefont {Karniadakis}}]{Fedosov:2011cf}%
  \BibitemOpen
  \bibfield  {author} {\bibinfo {author} {\bibfnamefont {D.~A.}\ \bibnamefont
  {Fedosov}}, \bibinfo {author} {\bibfnamefont {H.}~\bibnamefont {Lei}},
  \bibinfo {author} {\bibfnamefont {B.}~\bibnamefont {Caswell}}, \bibinfo
  {author} {\bibfnamefont {S.}~\bibnamefont {Suresh}}, \ and\ \bibinfo {author}
  {\bibfnamefont {G.~E.}\ \bibnamefont {Karniadakis}},\ }\bibfield  {title}
  {\enquote {\bibinfo {title} {{Multiscale modeling of red blood cell mechanics
  and blood flow in malaria}},}\ }\href@noop {} {\bibfield  {journal} {\bibinfo
   {journal} {PLoS Comp. Biol.}\ }\textbf {\bibinfo {volume} {7}},\ \bibinfo
  {pages} {e1002270} (\bibinfo {year} {2011})}\BibitemShut {NoStop}%
\bibitem [{\citenamefont {Yazdani}\ and\ \citenamefont
  {Bagchi}(2011)}]{Yazdani2011PRE}%
  \BibitemOpen
  \bibfield  {author} {\bibinfo {author} {\bibfnamefont {A.~Z.~K.}\
  \bibnamefont {Yazdani}}\ and\ \bibinfo {author} {\bibfnamefont
  {P.}~\bibnamefont {Bagchi}},\ }\bibfield  {title} {\enquote {\bibinfo {title}
  {Phase diagram and breathing dynamics of a single red blood cell and a
  biconcave capsule in dilute shear flow},}\ }\href {\doibase
  10.1103/PhysRevE.84.026314} {\bibfield  {journal} {\bibinfo  {journal} {Phys.
  Rev. E}\ }\textbf {\bibinfo {volume} {84}},\ \bibinfo {pages} {026314}
  (\bibinfo {year} {2011})}\BibitemShut {NoStop}%
\bibitem [{\citenamefont {Sinha}\ and\ \citenamefont
  {Graham}(2015)}]{Sinha:2015wt}%
  \BibitemOpen
  \bibfield  {author} {\bibinfo {author} {\bibfnamefont {K.}~\bibnamefont
  {Sinha}}\ and\ \bibinfo {author} {\bibfnamefont {M.~D.}\ \bibnamefont
  {Graham}},\ }\bibfield  {title} {\enquote {\bibinfo {title} {{Dynamics of a
  single red blood cell in simple shear flow}},}\ }\href@noop {} {\bibfield
  {journal} {\bibinfo  {journal} {Phys. Rev. E}\ }\textbf {\bibinfo {volume}
  {92}},\ \bibinfo {pages} {042710} (\bibinfo {year} {2015})}\BibitemShut
  {NoStop}%
\bibitem [{\citenamefont {Byun}\ \emph {et~al.}(2012)\citenamefont {Byun},
  \citenamefont {Hillman}, \citenamefont {Higgins}, \citenamefont {Diez-Silva},
  \citenamefont {Peng}, \citenamefont {Dao}, \citenamefont {Dasari},
  \citenamefont {Suresh},\ and\ \citenamefont {Park}}]{BYUN20124130}%
  \BibitemOpen
  \bibfield  {author} {\bibinfo {author} {\bibfnamefont {H.}~\bibnamefont
  {Byun}}, \bibinfo {author} {\bibfnamefont {T.~R.}\ \bibnamefont {Hillman}},
  \bibinfo {author} {\bibfnamefont {J.~M.}\ \bibnamefont {Higgins}}, \bibinfo
  {author} {\bibfnamefont {M.}~\bibnamefont {Diez-Silva}}, \bibinfo {author}
  {\bibfnamefont {Z.}~\bibnamefont {Peng}}, \bibinfo {author} {\bibfnamefont
  {M.}~\bibnamefont {Dao}}, \bibinfo {author} {\bibfnamefont {R.~R.}\
  \bibnamefont {Dasari}}, \bibinfo {author} {\bibfnamefont {S.}~\bibnamefont
  {Suresh}}, \ and\ \bibinfo {author} {\bibfnamefont {Y.}~\bibnamefont
  {Park}},\ }\bibfield  {title} {\enquote {\bibinfo {title} {Optical
  measurement of biomechanical properties of individual erythrocytes from a
  sickle cell patient},}\ }\href {\doibase
  https://doi.org/10.1016/j.actbio.2012.07.011} {\bibfield  {journal} {\bibinfo
   {journal} {Acta Biomater.}\ }\textbf {\bibinfo {volume} {8}},\ \bibinfo
  {pages} {4130--4138} (\bibinfo {year} {2012})}\BibitemShut {NoStop}%
\bibitem [{\citenamefont {Messer}\ and\ \citenamefont
  {Harris}(1970)}]{messer1970}%
  \BibitemOpen
  \bibfield  {author} {\bibinfo {author} {\bibfnamefont {M.~J.}\ \bibnamefont
  {Messer}}\ and\ \bibinfo {author} {\bibfnamefont {J.~W.}\ \bibnamefont
  {Harris}},\ }\bibfield  {title} {\enquote {\bibinfo {title} {Filtration
  characteristics of sickle cells: Rates of alteration of filterability after
  deoxygenation and reoxygenation, and correlations with sickling and
  unsickling},}\ }\href {\doibase 10.5555/uri:pii:0022214370902404} {\bibfield
  {journal} {\bibinfo  {journal} {J. Lab. Clin. Med.}\ }\textbf {\bibinfo
  {volume} {76}},\ \bibinfo {pages} {537--547} (\bibinfo {year}
  {1970})}\BibitemShut {NoStop}%
\bibitem [{\citenamefont {Nash}\ \emph {et~al.}(1984)\citenamefont {Nash},
  \citenamefont {Johnson},\ and\ \citenamefont {Meiselman}}]{Nash73}%
  \BibitemOpen
  \bibfield  {author} {\bibinfo {author} {\bibfnamefont {G.~B.}\ \bibnamefont
  {Nash}}, \bibinfo {author} {\bibfnamefont {C.~S.}\ \bibnamefont {Johnson}}, \
  and\ \bibinfo {author} {\bibfnamefont {H.~J.}\ \bibnamefont {Meiselman}},\
  }\bibfield  {title} {\enquote {\bibinfo {title} {Mechanical properties of
  oxygenated red blood cells in sickle cell ({HbSS}) disease},}\ }\href
  {http://www.bloodjournal.org/content/63/1/73} {\bibfield  {journal} {\bibinfo
   {journal} {Blood}\ }\textbf {\bibinfo {volume} {63}},\ \bibinfo {pages}
  {73--82} (\bibinfo {year} {1984})}\BibitemShut {NoStop}%
\bibitem [{\citenamefont {Evans}\ and\ \citenamefont
  {Mohandas}(1987)}]{Evans1443}%
  \BibitemOpen
  \bibfield  {author} {\bibinfo {author} {\bibfnamefont {E.~A.}\ \bibnamefont
  {Evans}}\ and\ \bibinfo {author} {\bibfnamefont {N.}~\bibnamefont
  {Mohandas}},\ }\bibfield  {title} {\enquote {\bibinfo {title}
  {Membrane-associated sickle hemoglobin: a major determinant of sickle
  erythrocyte rigidity},}\ }\href@noop {} {\bibfield  {journal} {\bibinfo
  {journal} {Blood}\ }\textbf {\bibinfo {volume} {70}},\ \bibinfo {pages}
  {1443--1449} (\bibinfo {year} {1987})}\BibitemShut {NoStop}%
\bibitem [{\citenamefont {Bertles}\ and\ \citenamefont
  {D{\"o}bler}(1969)}]{BERTLES884}%
  \BibitemOpen
  \bibfield  {author} {\bibinfo {author} {\bibfnamefont {J.~F.}\ \bibnamefont
  {Bertles}}\ and\ \bibinfo {author} {\bibfnamefont {J.}~\bibnamefont
  {D{\"o}bler}},\ }\bibfield  {title} {\enquote {\bibinfo {title} {Reversible
  and irreversible sickling: a distinction by electron microscopy},}\ }\href
  {http://www.bloodjournal.org/content/33/6/884} {\bibfield  {journal}
  {\bibinfo  {journal} {Blood}\ }\textbf {\bibinfo {volume} {33}},\ \bibinfo
  {pages} {884--898} (\bibinfo {year} {1969})}\BibitemShut {NoStop}%
\bibitem [{\citenamefont {Jeffery}(1922)}]{Jeffery:1922wb}%
  \BibitemOpen
  \bibfield  {author} {\bibinfo {author} {\bibfnamefont {G.~B.}\ \bibnamefont
  {Jeffery}},\ }\bibfield  {title} {\enquote {\bibinfo {title} {{The motion of
  ellipsoidal particles immersed in a viscous fluid}},}\ }\href@noop {}
  {\bibfield  {journal} {\bibinfo  {journal} {Proc. Roy. Soc. Lond. A}\
  }\textbf {\bibinfo {volume} {102}},\ \bibinfo {pages} {161--179} (\bibinfo
  {year} {1922})}\BibitemShut {NoStop}%
\bibitem [{\citenamefont {Bretherton}(1962)}]{bretherton_1962}%
  \BibitemOpen
  \bibfield  {author} {\bibinfo {author} {\bibfnamefont {F.~P.}\ \bibnamefont
  {Bretherton}},\ }\bibfield  {title} {\enquote {\bibinfo {title} {The motion
  of rigid particles in a shear flow at low {R}eynolds number},}\ }\href@noop
  {} {\bibfield  {journal} {\bibinfo  {journal} {J. Fluid Mech.}\ }\textbf
  {\bibinfo {volume} {14}},\ \bibinfo {pages} {284--304} (\bibinfo {year}
  {1962})}\BibitemShut {NoStop}%
\bibitem [{\citenamefont {Hinch}\ and\ \citenamefont
  {Leal}(1979)}]{hinch_leal_1979}%
  \BibitemOpen
  \bibfield  {author} {\bibinfo {author} {\bibfnamefont {E.~J.}\ \bibnamefont
  {Hinch}}\ and\ \bibinfo {author} {\bibfnamefont {L.~G.}\ \bibnamefont
  {Leal}},\ }\bibfield  {title} {\enquote {\bibinfo {title} {{Rotation of small
  non-axisymmetric particles in a simple shear flow}},}\ }\href@noop {}
  {\bibfield  {journal} {\bibinfo  {journal} {J. Fluid Mech.}\ }\textbf
  {\bibinfo {volume} {92}},\ \bibinfo {pages} {591--607} (\bibinfo {year}
  {1979})}\BibitemShut {NoStop}%
\bibitem [{\citenamefont {Yarin}\ \emph {et~al.}(1997)\citenamefont {Yarin},
  \citenamefont {Gottlieb},\ and\ \citenamefont {Roisman}}]{Yarin:1997cw}%
  \BibitemOpen
  \bibfield  {author} {\bibinfo {author} {\bibfnamefont {A.~L.}\ \bibnamefont
  {Yarin}}, \bibinfo {author} {\bibfnamefont {O.}~\bibnamefont {Gottlieb}}, \
  and\ \bibinfo {author} {\bibfnamefont {I.~V.}\ \bibnamefont {Roisman}},\
  }\bibfield  {title} {\enquote {\bibinfo {title} {{Chaotic rotation of
  triaxial ellipsoids in simple shear flow}},}\ }\href@noop {} {\bibfield
  {journal} {\bibinfo  {journal} {J. Fluid Mech.}\ }\textbf {\bibinfo {volume}
  {340}},\ \bibinfo {pages} {83--100} (\bibinfo {year} {1997})}\BibitemShut
  {NoStop}%
\bibitem [{\citenamefont {Dupont}\ \emph {et~al.}(2013)\citenamefont {Dupont},
  \citenamefont {Salsac},\ and\ \citenamefont {Barthes-Biesel}}]{Dupont2013}%
  \BibitemOpen
  \bibfield  {author} {\bibinfo {author} {\bibfnamefont {C.}~\bibnamefont
  {Dupont}}, \bibinfo {author} {\bibfnamefont {A.}~\bibnamefont {Salsac}}, \
  and\ \bibinfo {author} {\bibfnamefont {D.}~\bibnamefont {Barthes-Biesel}},\
  }\bibfield  {title} {\enquote {\bibinfo {title} {Off-plane motion of a
  prolate capsule in shear flow},}\ }\href@noop {} {\bibfield  {journal}
  {\bibinfo  {journal} {J. Fluid Mech.}\ }\textbf {\bibinfo {volume} {721}},\
  \bibinfo {pages} {180--198} (\bibinfo {year} {2013})}\BibitemShut {NoStop}%
\bibitem [{\citenamefont {Cordasco}\ and\ \citenamefont
  {Bagchi}(2013)}]{Cordasco:2013hb}%
  \BibitemOpen
  \bibfield  {author} {\bibinfo {author} {\bibfnamefont {D.}~\bibnamefont
  {Cordasco}}\ and\ \bibinfo {author} {\bibfnamefont {P.}~\bibnamefont
  {Bagchi}},\ }\bibfield  {title} {\enquote {\bibinfo {title} {{Orbital drift
  of capsules and red blood cells in shear flow}},}\ }\href@noop {} {\bibfield
  {journal} {\bibinfo  {journal} {Phys. Fluids}\ }\textbf {\bibinfo {volume}
  {25}},\ \bibinfo {pages} {091902} (\bibinfo {year} {2013})}\BibitemShut
  {NoStop}%
\bibitem [{\citenamefont {Mao}\ and\ \citenamefont
  {Alexeev}(2014)}]{mao_alexeev_2014}%
  \BibitemOpen
  \bibfield  {author} {\bibinfo {author} {\bibfnamefont {W.}~\bibnamefont
  {Mao}}\ and\ \bibinfo {author} {\bibfnamefont {A.}~\bibnamefont {Alexeev}},\
  }\bibfield  {title} {\enquote {\bibinfo {title} {Motion of spheroid particles
  in shear flow with inertia},}\ }\href {\doibase 10.1017/jfm.2014.224}
  {\bibfield  {journal} {\bibinfo  {journal} {J. Fluid Mech.}\ }\textbf
  {\bibinfo {volume} {749}},\ \bibinfo {pages} {145--166} (\bibinfo {year}
  {2014})}\BibitemShut {NoStop}%
\bibitem [{\citenamefont {Lundell}\ and\ \citenamefont
  {Carlsson}(2010)}]{Lundell2010}%
  \BibitemOpen
  \bibfield  {author} {\bibinfo {author} {\bibfnamefont {F.}~\bibnamefont
  {Lundell}}\ and\ \bibinfo {author} {\bibfnamefont {A.}~\bibnamefont
  {Carlsson}},\ }\bibfield  {title} {\enquote {\bibinfo {title} {Heavy
  ellipsoids in creeping shear flow: Transitions of the particle rotation rate
  and orbit shape},}\ }\href {\doibase 10.1103/PhysRevE.81.016323} {\bibfield
  {journal} {\bibinfo  {journal} {Phys. Rev. E}\ }\textbf {\bibinfo {volume}
  {81}},\ \bibinfo {pages} {016323} (\bibinfo {year} {2010})}\BibitemShut
  {NoStop}%
\bibitem [{\citenamefont {Wang}\ \emph {et~al.}(2012)\citenamefont {Wang},
  \citenamefont {Tozzi}, \citenamefont {Graham},\ and\ \citenamefont
  {Klingenberg}}]{Wang:2012ji}%
  \BibitemOpen
  \bibfield  {author} {\bibinfo {author} {\bibfnamefont {J.}~\bibnamefont
  {Wang}}, \bibinfo {author} {\bibfnamefont {E.~J.}\ \bibnamefont {Tozzi}},
  \bibinfo {author} {\bibfnamefont {M.~D.}\ \bibnamefont {Graham}}, \ and\
  \bibinfo {author} {\bibfnamefont {D.~J.}\ \bibnamefont {Klingenberg}},\
  }\bibfield  {title} {\enquote {\bibinfo {title} {{Flipping, scooping, and
  spinning: Drift of rigid curved nonchiral fibers in simple shear flow}},}\
  }\href@noop {} {\bibfield  {journal} {\bibinfo  {journal} {Phys. Fluids}\
  }\textbf {\bibinfo {volume} {24}},\ \bibinfo {pages} {123304} (\bibinfo
  {year} {2012})}\BibitemShut {NoStop}%
\bibitem [{\citenamefont {Crowdy}(2016)}]{Crowdy2016}%
  \BibitemOpen
  \bibfield  {author} {\bibinfo {author} {\bibfnamefont {D.}~\bibnamefont
  {Crowdy}},\ }\bibfield  {title} {\enquote {\bibinfo {title} {Flipping and
  scooping of curved 2{D} rigid fibers in simple shear: The {J}effery
  equations},}\ }\href {\doibase 10.1063/1.4948776} {\bibfield  {journal}
  {\bibinfo  {journal} {Phys. Fluids}\ }\textbf {\bibinfo {volume} {28}},\
  \bibinfo {pages} {053105} (\bibinfo {year} {2016})}\BibitemShut {NoStop}%
\bibitem [{\citenamefont {Dutta}\ and\ \citenamefont
  {Graham}(2017)}]{dutta2017}%
  \BibitemOpen
  \bibfield  {author} {\bibinfo {author} {\bibfnamefont {S.}~\bibnamefont
  {Dutta}}\ and\ \bibinfo {author} {\bibfnamefont {M.~D.}\ \bibnamefont
  {Graham}},\ }\bibfield  {title} {\enquote {\bibinfo {title} {Dynamics of
  {M}iura-patterned foldable sheets in shear flow},}\ }\href {\doibase
  10.1039/C6SM02113A} {\bibfield  {journal} {\bibinfo  {journal} {Soft Matter}\
  }\textbf {\bibinfo {volume} {13}},\ \bibinfo {pages} {2620--2633} (\bibinfo
  {year} {2017})}\BibitemShut {NoStop}%
\bibitem [{\citenamefont {Skalak}\ \emph {et~al.}(1973)\citenamefont {Skalak},
  \citenamefont {Tozeren}, \citenamefont {Zarda},\ and\ \citenamefont
  {Chien}}]{SKALAK:1973tp}%
  \BibitemOpen
  \bibfield  {author} {\bibinfo {author} {\bibfnamefont {R.}~\bibnamefont
  {Skalak}}, \bibinfo {author} {\bibfnamefont {A.}~\bibnamefont {Tozeren}},
  \bibinfo {author} {\bibfnamefont {R.~P.}\ \bibnamefont {Zarda}}, \ and\
  \bibinfo {author} {\bibfnamefont {S.}~\bibnamefont {Chien}},\ }\bibfield
  {title} {\enquote {\bibinfo {title} {{Strain energy function of red
  blood-cell membranes}},}\ }\href@noop {} {\bibfield  {journal} {\bibinfo
  {journal} {Biophys. J.}\ }\textbf {\bibinfo {volume} {13}},\ \bibinfo {pages}
  {245--280} (\bibinfo {year} {1973})}\BibitemShut {NoStop}%
\bibitem [{\citenamefont {P.~Mills}\ \emph {et~al.}(2004)\citenamefont
  {P.~Mills}, \citenamefont {Qie}, \citenamefont {Dao}, \citenamefont {Lim},\
  and\ \citenamefont {Suresh}}]{Mills2004}%
  \BibitemOpen
  \bibfield  {author} {\bibinfo {author} {\bibfnamefont {J.}~\bibnamefont
  {P.~Mills}}, \bibinfo {author} {\bibfnamefont {L.}~\bibnamefont {Qie}},
  \bibinfo {author} {\bibfnamefont {M.}~\bibnamefont {Dao}}, \bibinfo {author}
  {\bibfnamefont {C.~T.}\ \bibnamefont {Lim}}, \ and\ \bibinfo {author}
  {\bibfnamefont {S.}~\bibnamefont {Suresh}},\ }\bibfield  {title} {\enquote
  {\bibinfo {title} {Nonlinear elastic and viscoelastic deformation of the
  human red blood cell with optical tweezers},}\ }\href@noop {} {\bibfield
  {journal} {\bibinfo  {journal} {Mech. Chem. Biosyst.}\ }\textbf {\bibinfo
  {volume} {1}},\ \bibinfo {pages} {169--180} (\bibinfo {year}
  {2004})}\BibitemShut {NoStop}%
\bibitem [{\citenamefont {Lipowsky}(2013)}]{Lipowsky2013}%
  \BibitemOpen
  \bibfield  {author} {\bibinfo {author} {\bibfnamefont {H.~H.}\ \bibnamefont
  {Lipowsky}},\ }\bibfield  {title} {\enquote {\bibinfo {title} {In vivo
  studies of blood rheology in the microcirculation in an in vitro world:
  {P}ast, present and future},}\ }\href@noop {} {\bibfield  {journal} {\bibinfo
   {journal} {Biorheology}\ }\textbf {\bibinfo {volume} {50}},\ \bibinfo
  {pages} {3--16} (\bibinfo {year} {2013})}\BibitemShut {NoStop}%
\bibitem [{\citenamefont {Laogun}\ \emph {et~al.}(1980)\citenamefont {Laogun},
  \citenamefont {Ajayi}, \citenamefont {Osamo},\ and\ \citenamefont
  {Okafor}}]{Laogun1980}%
  \BibitemOpen
  \bibfield  {author} {\bibinfo {author} {\bibfnamefont {A.~A.}\ \bibnamefont
  {Laogun}}, \bibinfo {author} {\bibfnamefont {N.~O.}\ \bibnamefont {Ajayi}},
  \bibinfo {author} {\bibfnamefont {N.~O.}\ \bibnamefont {Osamo}}, \ and\
  \bibinfo {author} {\bibfnamefont {L.~A.}\ \bibnamefont {Okafor}},\ }\bibfield
   {title} {\enquote {\bibinfo {title} {Plasma viscosity in sickle-cell
  anaemia},}\ }\href {http://stacks.iop.org/0143-0815/1/i=2/a=005} {\bibfield
  {journal} {\bibinfo  {journal} {Clin. Phys. Physiol. Meas.}\ }\textbf
  {\bibinfo {volume} {1}},\ \bibinfo {pages} {145} (\bibinfo {year}
  {1980})}\BibitemShut {NoStop}%
\bibitem [{\citenamefont {Evans}\ \emph {et~al.}(2008)\citenamefont {Evans},
  \citenamefont {Gratzer}, \citenamefont {Mohandas}, \citenamefont {Parker},\
  and\ \citenamefont {Sleep}}]{Evans2008}%
  \BibitemOpen
  \bibfield  {author} {\bibinfo {author} {\bibfnamefont {J.}~\bibnamefont
  {Evans}}, \bibinfo {author} {\bibfnamefont {W.}~\bibnamefont {Gratzer}},
  \bibinfo {author} {\bibfnamefont {N.}~\bibnamefont {Mohandas}}, \bibinfo
  {author} {\bibfnamefont {K.}~\bibnamefont {Parker}}, \ and\ \bibinfo {author}
  {\bibfnamefont {J.}~\bibnamefont {Sleep}},\ }\bibfield  {title} {\enquote
  {\bibinfo {title} {Fluctuations of the red blood cell membrane: relation to
  mechanical properties and lack of {ATP} dependence},}\ }\href {\doibase
  10.1529/biophysj.107.117952} {\bibfield  {journal} {\bibinfo  {journal}
  {Biophys. J.}\ }\textbf {\bibinfo {volume} {94}},\ \bibinfo {pages}
  {4134--4144} (\bibinfo {year} {2008})}\BibitemShut {NoStop}%
\bibitem [{\citenamefont {Betz}\ \emph {et~al.}(2009)\citenamefont {Betz},
  \citenamefont {Lenz}, \citenamefont {Joanny},\ and\ \citenamefont
  {Sykes}}]{Betz15320}%
  \BibitemOpen
  \bibfield  {author} {\bibinfo {author} {\bibfnamefont {T.}~\bibnamefont
  {Betz}}, \bibinfo {author} {\bibfnamefont {M.}~\bibnamefont {Lenz}}, \bibinfo
  {author} {\bibfnamefont {J.}~\bibnamefont {Joanny}}, \ and\ \bibinfo {author}
  {\bibfnamefont {C.}~\bibnamefont {Sykes}},\ }\bibfield  {title} {\enquote
  {\bibinfo {title} {A{TP}-dependent mechanics of red blood cells},}\ }\href
  {\doibase 10.1073/pnas.0904614106} {\bibfield  {journal} {\bibinfo  {journal}
  {Proc. Natl. Acad. Sci. USA}\ }\textbf {\bibinfo {volume} {106}},\ \bibinfo
  {pages} {15320--15325} (\bibinfo {year} {2009})}\BibitemShut {NoStop}%
\bibitem [{\citenamefont {Kumar}\ and\ \citenamefont
  {Graham}(2012)}]{Kumar:2012ev}%
  \BibitemOpen
  \bibfield  {author} {\bibinfo {author} {\bibfnamefont {A.}~\bibnamefont
  {Kumar}}\ and\ \bibinfo {author} {\bibfnamefont {M.~D.}\ \bibnamefont
  {Graham}},\ }\bibfield  {title} {\enquote {\bibinfo {title} {{Accelerated
  boundary integral method for multiphase flow in non-periodic geometries}},}\
  }\href@noop {} {\bibfield  {journal} {\bibinfo  {journal} {J. Comput. Phys.}\
  }\textbf {\bibinfo {volume} {231}},\ \bibinfo {pages} {6682--6713} (\bibinfo
  {year} {2012})}\BibitemShut {NoStop}%
\bibitem [{\citenamefont {Charrier}\ \emph {et~al.}(1989)\citenamefont
  {Charrier}, \citenamefont {Shrivastava},\ and\ \citenamefont
  {Wu}}]{charrier1989free}%
  \BibitemOpen
  \bibfield  {author} {\bibinfo {author} {\bibfnamefont {J.~M.}\ \bibnamefont
  {Charrier}}, \bibinfo {author} {\bibfnamefont {S.}~\bibnamefont
  {Shrivastava}}, \ and\ \bibinfo {author} {\bibfnamefont {R.}~\bibnamefont
  {Wu}},\ }\bibfield  {title} {\enquote {\bibinfo {title} {Free and constrained
  inflation of elastic membranes in relation to thermoforming --
  non-axisymmetric problems},}\ }\href@noop {} {\bibfield  {journal} {\bibinfo
  {journal} {J. Strain Anal. Eng. Des.}\ }\textbf {\bibinfo {volume} {24}},\
  \bibinfo {pages} {55--74} (\bibinfo {year} {1989})}\BibitemShut {NoStop}%
\bibitem [{\citenamefont {Meyer}\ \emph {et~al.}(2002)\citenamefont {Meyer},
  \citenamefont {Desbrun}, \citenamefont {Schroeder},\ and\ \citenamefont
  {Barr}}]{Meyer:2002vh}%
  \BibitemOpen
  \bibfield  {author} {\bibinfo {author} {\bibfnamefont {M.}~\bibnamefont
  {Meyer}}, \bibinfo {author} {\bibfnamefont {M.}~\bibnamefont {Desbrun}},
  \bibinfo {author} {\bibfnamefont {P.}~\bibnamefont {Schroeder}}, \ and\
  \bibinfo {author} {\bibfnamefont {A.~H.}\ \bibnamefont {Barr}},\ }\bibfield
  {title} {\enquote {\bibinfo {title} {{Discrete differential geometry
  operators for triangulated 2-manifolds}},}\ }\href@noop {} {\bibfield
  {journal} {\bibinfo  {journal} {Visualization and Mathematics}\ }\textbf
  {\bibinfo {volume} {3}},\ \bibinfo {pages} {34--57} (\bibinfo {year}
  {2002})}\BibitemShut {NoStop}%
\bibitem [{\citenamefont {Pozrikidis}(1992)}]{pozrikidis1992boundary}%
  \BibitemOpen
  \bibfield  {author} {\bibinfo {author} {\bibfnamefont {C.}~\bibnamefont
  {Pozrikidis}},\ }\href {https://books.google.com/books?id=qXt5bOqDEgQC}
  {\emph {\bibinfo {title} {Boundary integral and singularity methods for
  linearized viscous flow}}},\ Cambridge Texts in Applied Mathematics\
  (\bibinfo  {publisher} {Cambridge University Press},\ \bibinfo {year}
  {1992})\BibitemShut {NoStop}%
\end{thebibliography}
\end{document}